\documentclass{aastex} 

\usepackage{natbib}

\usepackage[usenames,dvipsnames]{color}

\begin{document}

\title{Disentangling the role of environmental processes in galaxy
  clusters}


\author{Jonathan D. Hern\'andez-Fern\'andez\altaffilmark{1},
  J. M. V\'ilchez\altaffilmark{1}, and
  J. Iglesias-P\'aramo\altaffilmark{1,2}}

\altaffiltext{1}{Instituto de Astrof\'isica de Andaluc\'ia, Glorieta
  de la Astronom\'ia s/n, 18008 Granada; jonatan@iaa.es}

\altaffiltext{2}{Centro Astron\'omico Hispano Alem\'an C/ Jes\'us
  Durb\'an Rem\'on, 2-2 04004 Almer\'ia}

\begin{abstract}

In this work we present the results of a novel approach devoted to
disentangle the role of the environmental processes affecting galaxies
in clusters.

This is based on the analysis of the $NUV-r'_{}$ distributions of a
large sample of star-forming galaxies in clusters spanning more than
four absolute magnitudes. The galaxies inhabit three distinct
environmental regions: virial regions, cluster infall regions and
field environment.

We have applied rigorous statistical tests in order to analyze both,
the complete $NUV-r'_{}$ distributions and their averages for three
different bins of $r'$-band galaxy luminosity down to
$M_{r'}$$\sim$-18, throughout the three environmental regions
considered.

We have identified the environmental processes that significantly
affect the star-forming galaxies in a given luminosity bin by using
criteria based on the characteristics of these processes: their
typical time-scales, the regions where they operate and the galaxy
luminosity range for which their effects are more intense.

We have found that the high-luminosity \mbox{($M_{r'}$$\le$-20)}
star-forming galaxies do not show significant signs in their star
formation activity neither of being affected by the environment in the
last $\sim$10$^{8}$ yr nor of a sudden quenching in the last 1.5 Gyr.

The intermediate-luminosity \mbox{(-20$<$$M_{r'}$$\le$-19)} star-forming
galaxies appear to be affected by starvation in the virial regions and
by the harassment both, in the virial and infall regions.

Low-luminosity \mbox{(-19$<$$M_{r'}$$\le$-18.2)} star-forming galaxies seem
to be affected by the same environmental processes as
intermediate-luminosity star-forming galaxies in a stronger way, as it
would be expected for their lower luminosities.

\end{abstract}



\section{Introduction}

The influence of the environment on galaxies involves a rich variety
of processes that include the interactions of galaxies with other
components of the Universe: other galaxies, the intra-cluster medium
(ICM) or the cluster/group dark matter haloes (DMHs). As a result of
the different studies devoted to shed light on this key issue of the
Extragalactic Astronomy, a number of reviews have become available in
the recent years
\citep{Treu_et_al_2003,Poggianti_2006,Boselli&Gavazzi_2006,Mo_et_al_2010}. As
a reference, \citet{Treu_et_al_2003} split the environmental processes
into three groups:

\begin{itemize}

\item Galaxy-ICM interactions dominate the gas stripping processes,
  where the insterstellar medium of galaxies is stripped via various
  mechanisms, including viscous and turbulent stripping, thermal
  evaporation and ram-pressure stripping
  \citep{Boselli&Gavazzi_2006}. Galaxy-ICM interactions can also
  trigger star formation through the compression of galactic gas
  clouds \citep{Dressler&Gunn_1983,Evrard_1991,Fujita_1998}. The ICM
  stripping of the hot halo gas results in a subsequent quenching of
  star formation \citep{Bekki_et_al_2002}.

\item Galaxy-cluster gravitational interactions can tidally compress
  the galaxy envelope of gas and increase the star formation rate
  \citep{Byrd&Valtonen_1990,Henriksen&Byrd_1996,Fujita_1998}. The
  galaxy starvation can be enhanced by the tidal interaction with the
  cluster DMH which contributes to the removal of the hot gas halo of
  the galaxy \citep{Bekki_et_al_2002}. The tidal truncation of the
  external galaxy regions due to the action of the cluster potential
  produces a late quenching of star formation along a few Gyr if the
  galaxy reservoir is removed. However, their effects are more clearly
  observed in the structural changes of the mass profile
  \citep{Merritt_1984,Ghigna_et_al_1998,Natarajan_et_al_1998}.

\item Tidal galaxy-galaxy interactions dominate the galaxy mergers or
  strong galaxy encounters \citep{Icke_1985,Mihos_1995,Bekki_1998} and
  the galaxy harassment
  \citep{Moore_et_al_1996,Moore_et_al_1999,Moore_et_al_1998}.

\end{itemize}

These three groups of processes have associated different spatial/time
scales and characteristics which we describe below:

\begin{itemize}

\item Time-scales. We can distinguish between long and
  short-time-scale processes. Following the scheme proposed by
  Poggianti (2006), the long time-scale processes embrace the gas
  stripping processes and the galaxy mergers or strong galaxy
  interactions and the long-time-scale processes embrace the galaxy
  starvation \citep{Larson_et_al_1980,Bekki_et_al_2002} and the galaxy
  harassment \citep{Moore_et_al_1996}. The time-scale for suppression
  of star formation in gas stripping goes from $\sim$10$^{7}$ yr
  \citep{Abadi_et_al_1999} to 10$^{8}$ yr \citep{Quilis_et_al_2000}
  for a Milky Way-like galaxy. In galaxy mergers, both the encounter
  time and the duration of the starburst phase have a time-scale of
  the order of 10$^{7}$ yr
  \citep{Mihos&Hernquist_1994_ULSB,Mihos&Hernquist_1996}. Numerical
  simulations \citep{Bekki_et_al_2002} indicate that the starvation
  combining the interaction with the ICM and the cluster potential has
  a time-scale of several Gyr. Galaxy harassment takes place along
  several cluster crossing times, i.e. a few gigayears
  \citep[][Appendix A]{Boselli&Gavazzi_2006,Moore_et_al_1998}.

\item Spatial-scales. Also we can distinguish between long and short
  spatial-scale processes. The galaxy starvation and the galaxy
  harassment have spatial-scales of the order of virial radius or
  longer \citep{Treu_et_al_2003,Boselli&Gavazzi_2006}. In the gas
  stripping, the cold gas of galaxies interacts hydrodynamically with
  the ICM from its vicinity, though this occurs along the way crossing
  the ICM of the virial region, thus it is assumed to be a long
  spatial-scale process. The main small-scale environmental processes
  are the galaxy mergers or galaxy-galaxy tidal interactions which
  occur in a spatial scale that is of the order of the size of
  galaxies.

\item Environments. Certain environmental processes operate more
  efficiently or occur more frequently in certain types of
  environments or places. The starvation or the gas stripping operate
  preferably in the central regions of clusters, where the ICM drag
  force and the cluster tidal forces are enough to strip the gas
  components of the galaxy. The galaxy harassment rate $f_{h}$ scales
  with the luminous galaxy density $\rho_{gal}$ and mass $M_{\ast}$ as
  \mbox{$f_{h}$ $\propto$ $\rho_{gal}$$M_{\ast}^{2}$ $\propto$
    $\rho_{gal}$ $r^{2}$}. In the simple case where local density
  scales smoothly as $r^{-2}$, the harassment rate should be
  independent of the cluster radius \citep{Treu_et_al_2003}. The
  merging processes are most efficient when the relative velocities
  between the galaxies are low, thus they are expected to be more
  efficient in galaxy groups than in the inner regions of
  clusters. The galaxy groups infalling to clusters represent ideal
  places for an environmental-driven galaxy evolution in the frame of
  a "preprocessing" scenario \citep{Cortese_et_al_2004,Mihos_2004}.

\item Galaxy luminosity range. In addition to this, it has to be taken
  into account that the effect of the different processes depend on
  the galaxy luminosities, some of them are more efficient operating
  on bright/giant galaxies \citep[gas stripping,][]{Quilis_et_al_2000}
  and other processes affecting to dwarf and low-surface brightness
  galaxies \citep[galaxy harassment,][]{Moore_et_al_1999}.

\end{itemize}

These physical processes have been proposed/invoked in order to
explain numerous observational facts pointing out that the environment
in which a galaxy inhabits has a profound impact on its evolution in
terms of defining both its structural properties and its star
formation histories. Perhaps the two most remarkable observational
results are the morphology-density relation - the fraction of
early-type galaxies increases towards high-density environments -
\citep{Dressler_1980} and the star formation-density relation - the
fraction of star-forming galaxies as well as the intensity of the star
formation per galaxy decreases towards high-density environments -
\citep[][]{Lewis_et_al_2002,Gomez_et_al_2003,Rines_et_al_2005}.

The UV luminosity has revealed as a good proxy of the recent star
formation because it comes from the more short-lived stars
$\tau$$<$10$^{8}$ yr
\citep{Kennicutt_1998,Kauffmann_et_al_2007,Martin_et_al_2005}. Hence,
we can trace time variations of around 100 Myr in the star formation
history of a galaxy using the ultraviolet luminosity
\citep{Leitherer_et_al_1999} and/or UV-optical colors
\citep{Kaviraj_et_al_2007,Kaviraj_E+A}. The UV luminosity is strongly
linked to the FIR luminosity ($\sim$8-1000 $\mu$m), the UV radiation
emitted by the young stellar populations is strongly absorbed by dust
in the star-forming molecular clouds and reemitted through IR
photons. So, both the UV-to-FIR luminosity budget as a proxy of the
attenuation suffered by the young stellar population
\citep[see][]{Buat_et_al_2005} or the bolometric luminosity from the
young stellar populations $L_{BOL}$=$L_{UV}$+$L_{FIR}$ as a proxy of
the star formation rate
\citep{Wang&Heckman_1996,Iglesias-Paramo_et_al_2006} also give
important insights on the star formation activity of galaxies.

On the other hand, assuming that field galaxies are the galaxies less
affected by the influence of the environment, we can take a sample of
field galaxies as a fiducial sample (approximated) free from
interactions i.e. in absence of environmental influence. Therefore, if
the distribution of a UV-optical color of a star-forming galaxy sample
\citep[i.e. those with blue UV-optical colors,
  e.g.][]{Haines_et_al_2008,Chilingarian&Zolotukhin_2011} is
statistically different from the corresponding distribution for a
field star-forming galaxy sample (at the same cosmic time), we can
conclude those galaxies are suffering a significant change in their
star formation level in a time-scale of around 10$^{8}$ yr or
longer. Assuming this fact, we develop a original approach which
permit us assign the intensity of those environmental processes
affecting the galaxies depending on their luminosity by means of the
analysis of the variations of the distributions of a UV-optical color
of star-forming galaxies throughout different environmental
regions. We choose the UV-optical color $NUV-r'_{}$ because these two
spectral bands trace two different time-scales of the star formation
history; the NUV-band traces a $\tau$$\sim$10$^{8}$ yr and the
$r'$-band a more long time-scale $\tau$$\sim$10$^{9}$-10$^{10}$ yr
\citep{Kennicutt_1998,Martin_et_al_2005}. Because of this, $NUV-r'_{}$
shows a tight correlation with the ratio of the recent star formation
over the past-averaged star formation \citep{Salim_et_al_2005}.

The remainder of the paper is organized as follows: in section
\ref{sec:gal_samp}, we describe the photometric data and the sample of
galaxies in clusters used in this work. In section
\ref{sec:NUVr_env_trend}, we present the main results derived from the
$NUV-r'_{}$ distributions of our sample of galaxies. Section
\ref{sec:discussion} presents the discussion of the implications
resulting from our analysis. In section \ref{sec:summ&conclu}, we
summarize the main results and conclusions of this work.


\section{The sample of cluster galaxies}
\label{sec:gal_samp}

This work is based on data compiled from the SDSS-DR6
\citep{Adelman-McCarthy_et_al_2008} and GALEX-AIS
\citep{Martin_et_al_2005} for a sample of cluster galaxies extensively
described in \citet[][hereafter Paper
  I]{Hernandez-Fernandez_et_al_2011_GC}. \textcolor{black}{This sample
  consists on a total of $\sim$5000 galaxies from 16 nearby clusters
  \mbox{($z$$<$0.05)} showing a rich variety in their characteristics,
  from poor to rich clusters with cluster velocity dispersions between
  \mbox{200$\lesssim$$\left ( \frac{\sigma_{c}}{\rm{km~s^{-1}}} \right
    )$$\lesssim$800}. The basic properties of these clusters are
  listed in Table \ref{CS}. The galaxies span a luminosity range
  containing the classical luminosity boundary between giant and dwarf
  galaxies \mbox{$M_B$$\sim$-18}. The selection of the clusters was
  constrained by the condition that they were covered by SDSS-DR6 and
  GALEX-AIS throughout sky areas corresponding to physical sizes of
  several virial radii, i.e. several Mpc. This condition ensures that
  our sample of galaxies probes different environments from the
  central regions of rich clusters to the low-density field, including
  the galaxy structures around clusters as groups, sheets and
  filaments.}  The Main Galaxy Sample of SDSS
\citep{Strauss_et_al_2002} consists \textcolor{black}{of those galaxies
  with $r'$-band petrosian magnitudes $r'$$<$17.77 and $r'$-band
  petrosian half-light surface brightnesses $\mu$$<$24.5 mag
  arcsec$^{-2}$, retrieved from the imaging survey which has a 95\%
  completeness limit for stars of $r'_{}$=22.2
  \citep{SDSS_EDR}}. GALEX has performed the first space UV imaging
all-sky survey in two bands (\mbox{FUV:1350-1750 \AA} and
\mbox{NUV:1750-2750 \AA}) down to a Galactic-extincted magnitude of
$m_{AB}$=20.5. The broad-band photometry from both SDSS-DR6 and
GALEX-AIS is carefully chosen to retrieve the total integrated flux
for each galaxy in the selected band. For the SDSS photometry, we
choose the composite-model magnitude, which is a weighted linear
combination of the fluxes extracted with a \mbox{\it De Vaucouleurs}
profile and an exponential radial profile fit to the surface
brightness profile of each galaxy \citep{SDSS_II}. For the GALEX
photometry, we choose the elliptical aperture photometry
\citep[\texttt{MAG\_AUTO} option in SExtractor code,
][]{Bertin&Arnouts_1996} in order to include the total integrated UV
flux for each galaxy source.

\begin{table}[p]
\caption{Main properties of the cluster sample.}
\begin{center}
\resizebox{!}{7cm}{ 
\begin{tabular}{c        c              c            c           c            c          c          c             c           }
\hline
ID$_{NED}$     & $\alpha$(J2000) & $\delta$(J2000) & $z_{med}$  & $\sigma_{c}$& $r_{200}$   & $n_{tot}$ & $\theta_{tot}$& log(L$_X$) \\
                       & deg          & deg        &           & km s$^{-1}$ & Mpc       &          &  deg         & L$_{\odot}$ \\
 (1)                   & (2)          & (3)        & (4)       & (5)        & (6)       &  (7)     &  (8)         &  (9)     \\
\hline
 UGCl 141              & 138.499      & 30.2094    & 0.0228    & 501.8      & 1.21      &  413     &  4.159       &  42.12  \\
 WBL 245               & 149.120      & 20.5119    & 0.0255    &  86.7      & 0.20      &  88      &  3.720       &  ...    \\
 UGCl 148 NED01        & 142.366      & 30.2139    & 0.0263    & 316.7      & 0.76      &  354     &  3.606       &  ...    \\
 ABELL 2199            & 247.154      & 39.5244    & 0.0303    & 756.2      & 1.83      &  1104    &  3.125       &  44.85  \\
 WBL 213               & 139.283      & 20.0403    & 0.0290    & 537.1      & 1.29      &  548     &  3.266       &  $\le$41.9  \\
 WBL 514               & 218.504      & 3.78111    & 0.0291    & 633.7      & 1.52      &  580     &  3.257       &  43.18  \\
 WBL 210               & 139.025      & 17.7242    & 0.0287    & 433.3      & 1.06      &  402     &  3.298       &  43.22  \\
 WBL 234               & 145.602      & 4.27111    & 0.0291    & 243.6      & 0.58      &  87      &  3.262       &  ...    \\
 WBL 205               & 137.387      & 20.4464    & 0.0288    & 679.8      & 1.60      &  527     &  3.289       &  ...    \\
 UGCl 393              & 244.500      & 35.1000    & 0.0314    & 637.9      & 1.52      &  529     &  3.016       &  43.60  \\
 UGCl 391              & 243.352      & 37.1575    & 0.0330    & 407.0      & 0.97      &  637     &  2.874       &  ...    \\
 B2 1621+38:[MLO2002]  & 245.583      & 37.9611    & 0.0311    & 607.3      & 1.46      &  1053    &  3.046       &  43.19  \\
 UGCl 271              & 188.546      & 47.8911    & 0.0305    & 323.2      & 0.72      &  181     &  3.104       &  ...    \\
 ABELL 1185            & 167.699      & 28.6783    & 0.0328    & 789.3      & 1.90      &  754     &  2.894       &  43.58  \\
 ABELL 1213            & 169.121      & 29.2603    & 0.0469    & 565.7      & 1.35      &  305     &  2.021       &  43.77  \\
 UGCl 123 NED01        & 127.322      & 30.4828    & 0.0499    & 849.0      & 2.00      &  260     &  1.900       &  44.32  \\
\hline
\label{CS}
\end{tabular}} 
\end{center}
(1) NED identifier, (2) and (3) Celestial coordinates of cluster
center from NED webpage, (4) Cluster average redshift, (5) Cluster
velocity dispersion, (6) Radius 200, (7) No. of observed galaxies
associated to each cluster, (8) Half size of sky square region
retrieved for each cluster, computed assuming the Local Universe
approximation c$z$=H$D$, the small-angle approximation
$D_{P}$=$D$$\times$$\theta$[rad] and a projected radius $R_{P}$=7.1
Mpc (9) Bolometric X-ray luminosity from \citet{Mahdavi&Geller_2001}
except for WBL 213 \citep{Mahdavi_et_al_2000}.
\end{table}

\section{Results: UV-optical color distribution of cluster galaxies}
\label{sec:NUVr_env_trend}

The main goal of this work is to characterize the role of the
environment in the star formation properties of a sample of cluster
galaxies with active star formation. As a first step we select the
star forming galaxies from the total cluster galaxy sample previously
described. For this we apply the UV-optical color cut proposed for the
($NUV-r'_{}$) vs. ($u'-r'$) color-color diagram in Paper
I. \textcolor{black}{We assume a galaxy is a star-forming galaxy whether
  its colors fulfill the following prescription:}

\begin{displaymath}
{\rm STAR-FORMING:} \left\{ \begin{array}{ll}
{\it NUV-r'} < 4.9                & \textrm{, for } {\it u'-r'} < 2.175 \\
{\it NUV-r'} < -2({\it u'-r'}) + 9.25 & \textrm{, for } {\it u'-r'} > 2.175 
\label{eq:NUVr_ur_cut}
\end{array} \right.
\end{displaymath}

\begin{figure}[p]
\centering
\resizebox{1.00\hsize}{!}{\includegraphics{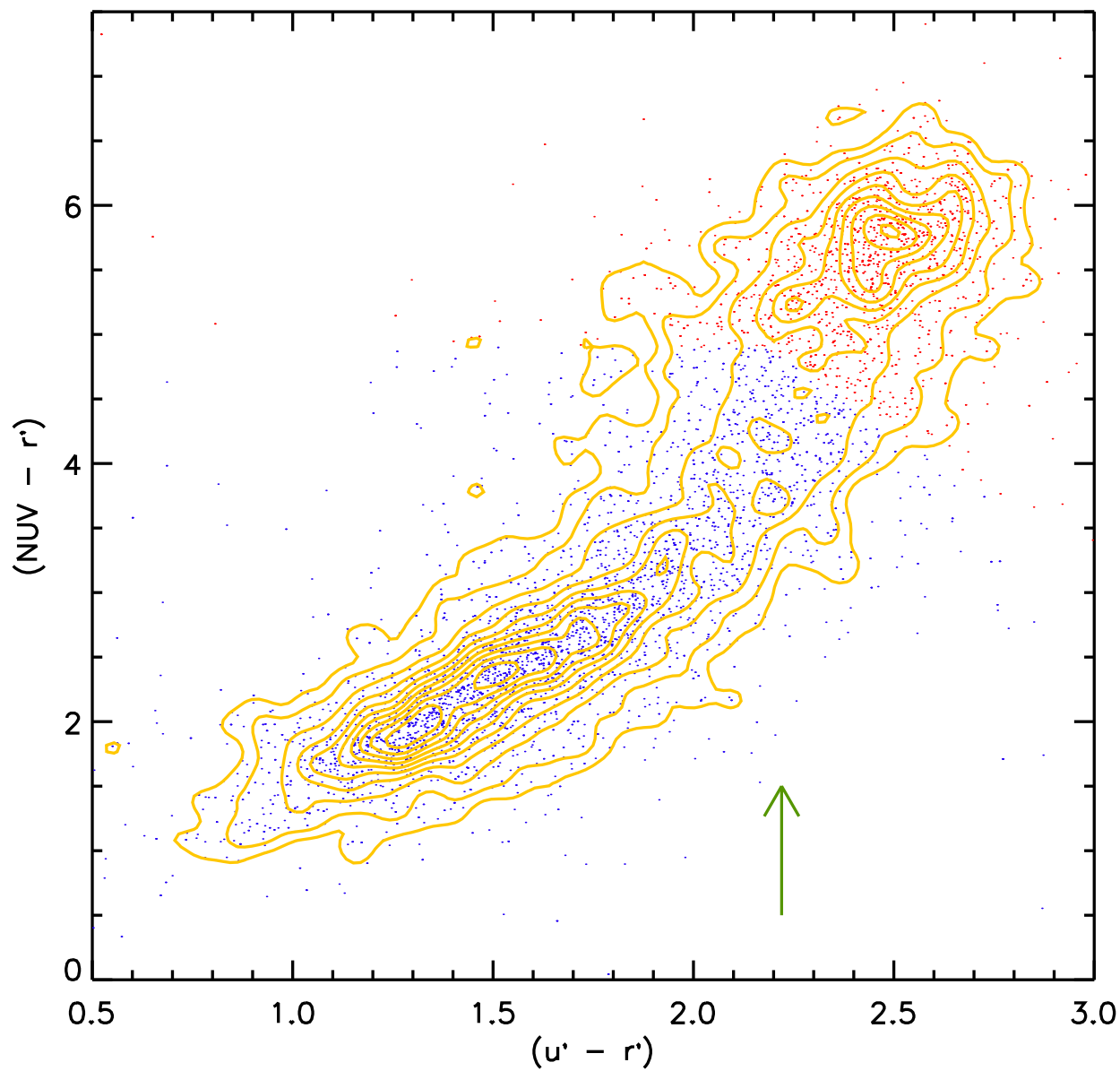}}
\caption{($NUV-r'_{}$) vs. ($u'-r'$) for the sample of galaxies. Yellow
  isocontours represents the isodensity contours of galaxies. Green
  dashed broken line is the UV-optical color-color cut proposed in
  Paper I. Green vertical arrow points out to the $u'-r'$ cut proposed
  by \citet{Strateva_et_al_2001}. Blue and red points represent,
  respectively, star-forming and passive galaxies under the
  prescription shown in Section \ref{eq:NUVr_ur_cut}.}
\label{NUVr_ur}
\end{figure}

\textcolor{black}{As can be seen in Figure \ref{NUVr_ur}, this boundary
  appears more appropriate to segregate star-forming galaxies from
  passive galaxies than the optical cut in the $u'-r'$ color proposed
  by \citet{Strateva_et_al_2001}.}

The next step is to define three regions where different physical
processes are expected to act \textcolor{black}{in a different way and
  intensity} due to the different environmental conditions: the
cluster virial regions, the cluster infall regions and the field
environment. In order to select galaxies from each environment, we
take advantage from their spatial ${\tilde r}$ and velocity ${\tilde
  s}$ variables through the phase diagram:

\begin{equation} 
({\tilde s},{\tilde r})\equiv \left( \left( \frac{{\rm c}(z-z_{med})}{\sigma_{c}} \right) , \left( \frac{R_{P}}{r_{200}} \right) \right)
\end{equation} 

being c the light speed, the $z$ galaxy redshift, $z_{med}$ cluster
redshift, $\sigma_{c}$ cluster velocity dispersion, $R_{P}$ physical
projected radius from cluster center and $r_{200}$ the virial radius.

In this phase diagram, galaxies in clusters are located inside the
region delimited by a bi-valued caustic ${\tilde s} = \pm {\cal
  A}$(${\tilde r}$) with a characteristic trumpet shape symmetrical
with respect to the ${\tilde r}$-axis \citep{Kaiser_1987}; outside the
caustic limits, the density of galaxies drops substantially. Galaxies
outside the caustics are background or foreground galaxies
\citep[e.g.][]{den_Hartog&Katgert_1996}. \citet{CAIRNS_I} numerically
compute the caustic ${\cal A}$(${\tilde r}$) for a sample of galaxy
clusters in the Local Universe (see their figure 4), using the method
proposed by \citet{Diaferio_1999}. This caustic curve can be
accurately approximated by the following terms:


\begin{displaymath}
{\cal A}(\tilde{r}) = \left\{ \begin{array}{ll}
2.5 - \frac{5}{7} \tilde{r} & \textrm{, for }{\tilde r}\le2_{\cdot}30  \\ 
1.0 - \frac{1}{16}\tilde{r} & \textrm{, for }{\tilde r}>2_{\cdot}30   
\end{array} \right.
\end{displaymath}

\textcolor{black}{Following \citet{CAIRNS_I}, we define the three
  different environments studied here in the following way:}


\begin{displaymath}
\begin{array}{rccl}
\textrm{Virial regions:}     & {\cal -A}({\tilde r})\le{\tilde s}\le{\cal +A}({\tilde r}) & \bigwedge & ~~~~~~{\tilde r}\le1  \\
\textrm{Infall regions:}     & {\cal -A}({\tilde r})\le{\tilde s}\le{\cal +A}({\tilde r}) & \bigwedge &   1<{\tilde r}\le5    \\ 
\textrm{Field environment:}  & {\tilde s}  <  {\cal -A}({\tilde r}) ~~~ \bigwedge ~~~ {\cal +A}({\tilde r})  <  {\tilde s}   & \bigvee   &   5<{\tilde r}
\end{array}
\end{displaymath}



Two galaxy clusters from Paper I, ABELL 2199 and WBL 514, present two
evident galaxy systems in their surroundings between $\sim$2$r_{200}$
and $\sim$3$r_{200}$ (see Figures 1 and 2 in Paper I). In the case of
WBL 514, this galaxy system corresponds to the cluster WBL 518 located
$\sim$2$^{\circ}$ west on the sky with coordinates
\mbox{($\alpha$,$\delta$)$_{J2000}$=}\mbox{( 14h 40m 43.1s , +03d 27m
  11s )} and a central redshift of $z$=0.027. The cluster ABELL 2199
has the cluster ABELL 2197 in their surroundings, located
$\sim$1.5$^{\circ}$ of projected distance, their coordinates are
\mbox{($\alpha$,$\delta$)$_{J2000}$=}\mbox{( 16h 28m 10.4s , +40d 54m
  26s )} and has a redshift of $z$=0.0308. These two galaxy systems,
WBL 518 and ABELL 2197, are not included in the set of clusters from
Paper I according to the constrain of being observed by the SDSS-DR6
and by the GALEX-AIS throughout sky regions corresponding to several
megaparsecs. The central regions of these galaxy systems can be only
inhabited by the typical galaxy population of the central regions of
galaxy clusters. In order to avoid the mix between galaxy population
from virial regions and from infall regions, we exclude the galaxy
subsample of these two clusters from the galaxy sample corresponding
to the infall regions. Specifically, we exclude from the infall
regions those galaxies with the following coordinates:

\begin{displaymath}
\begin{array}{r rcccr c rcccl }
{\rm ABELL~2197 :} & 40.4 & < & \alpha_{J2000}(^{\circ}) & < & 41.8 & \bigwedge  & 245.8  & < & \delta_{J2000}(^{\circ}) & < & 248.0  \\
   {\rm WBL~518 :} &  3.0 & < & \alpha_{J2000}(^{\circ}) & < &  4.2 & \bigwedge  & 219.8  & < & \delta_{J2000}(^{\circ}) & < & 220.75
\end{array}
\end{displaymath}



In Appendix, we show the results of the same analysis performed here
but including the central regions of these clusters as part of the
galaxy sample in the infall regions. 

\begin{figure}[p]
\centering
\resizebox{1.00\hsize}{!}{\includegraphics{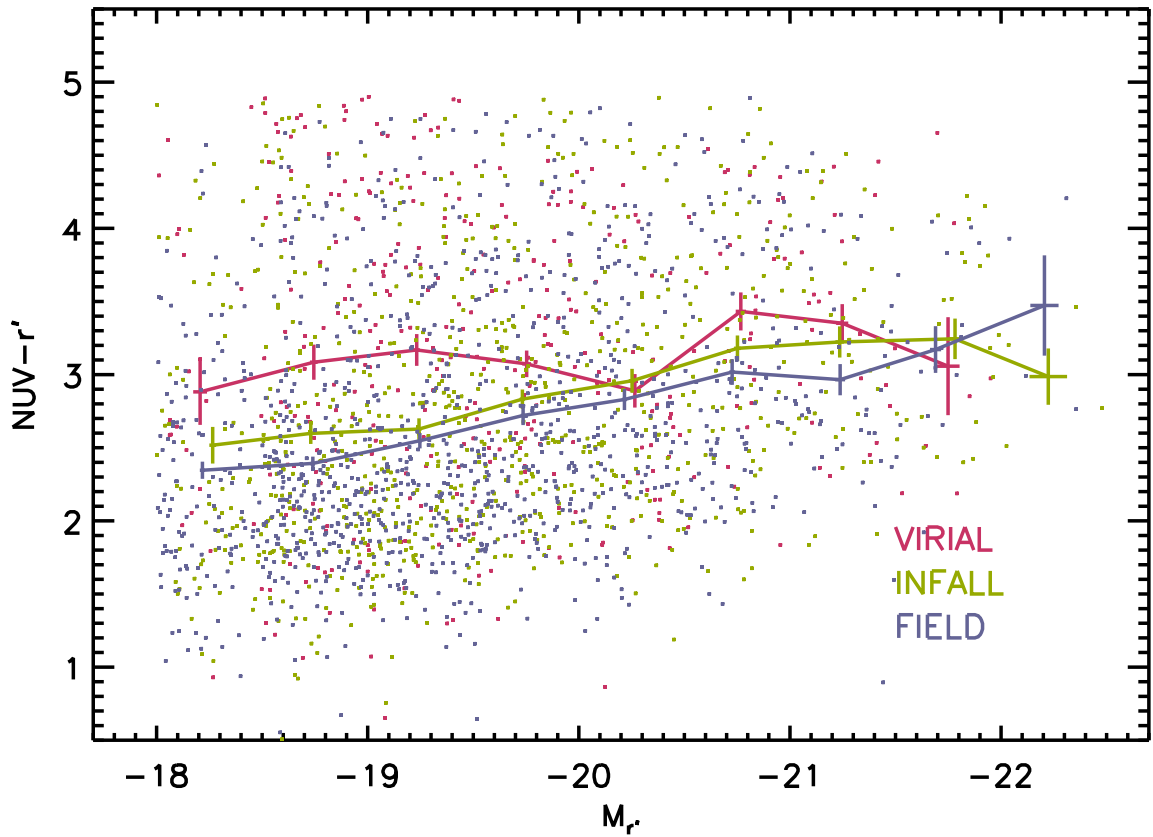}}
\caption{$NUV-r'_{}$ vs. $M_{r'}$ of star-forming galaxies in each of
  the defined environments: virial regions (red), infall regions
  (green) and field environment (blue). Each point corresponds to one
  galaxy from the environment identified by its color. The data points
  correspond to the mean of $NUV-r'_{}$ in each luminosity bin. The
  vertical and horizontal bars correspond to the bootstrap
  uncertainties in the mean of $NUV-r'_{}$ and in the mean of
  $M_{r'}$, respectively, in each luminosity bin (see text for
  details).}
\label{NUVr_Mr}
\end{figure}

Figure \ref{NUVr_Mr} shows the distribution and the average trend of
the $NUV-r'_{}$ color for our sample of star-forming galaxies along
their $r'$-band absolute magnitude $M_{r'}$ in the three different
environments. In order to be complete in each luminosity bin, we
select only those star-forming galaxies fulfilling the two following
conditions: ({\it i}) galaxies whose $r'$-band composite-model
absolute magnitude, $M_{r'}$, is inside the limits (lower limit -
$M^{lbin}_{r'}$; upper limit - $M^{ubin}_{r'}$) of each luminosity
bin,

\begin{equation}
M^{lbin}_{r'} < M_{r'} \le M^{ubin}_{r'}
\label{eq:g_comp_rule}
\end{equation} 

and ({\it ii}) galaxies from those clusters for which their
completeness limit of the $r'$-band composite-model absolute magnitude
$M^{clim}_{r'}$ is fainter than the upper limit of each luminosity bin
$M^{ubin}_{r'}$,

\begin{equation}
M^{ubin}_{r'} \le M^{clim}_{r'}
\label{eq:c_comp_rule}
\end{equation}

\begin{figure}[p]
\centering
\resizebox{1.00\hsize}{!}{\includegraphics{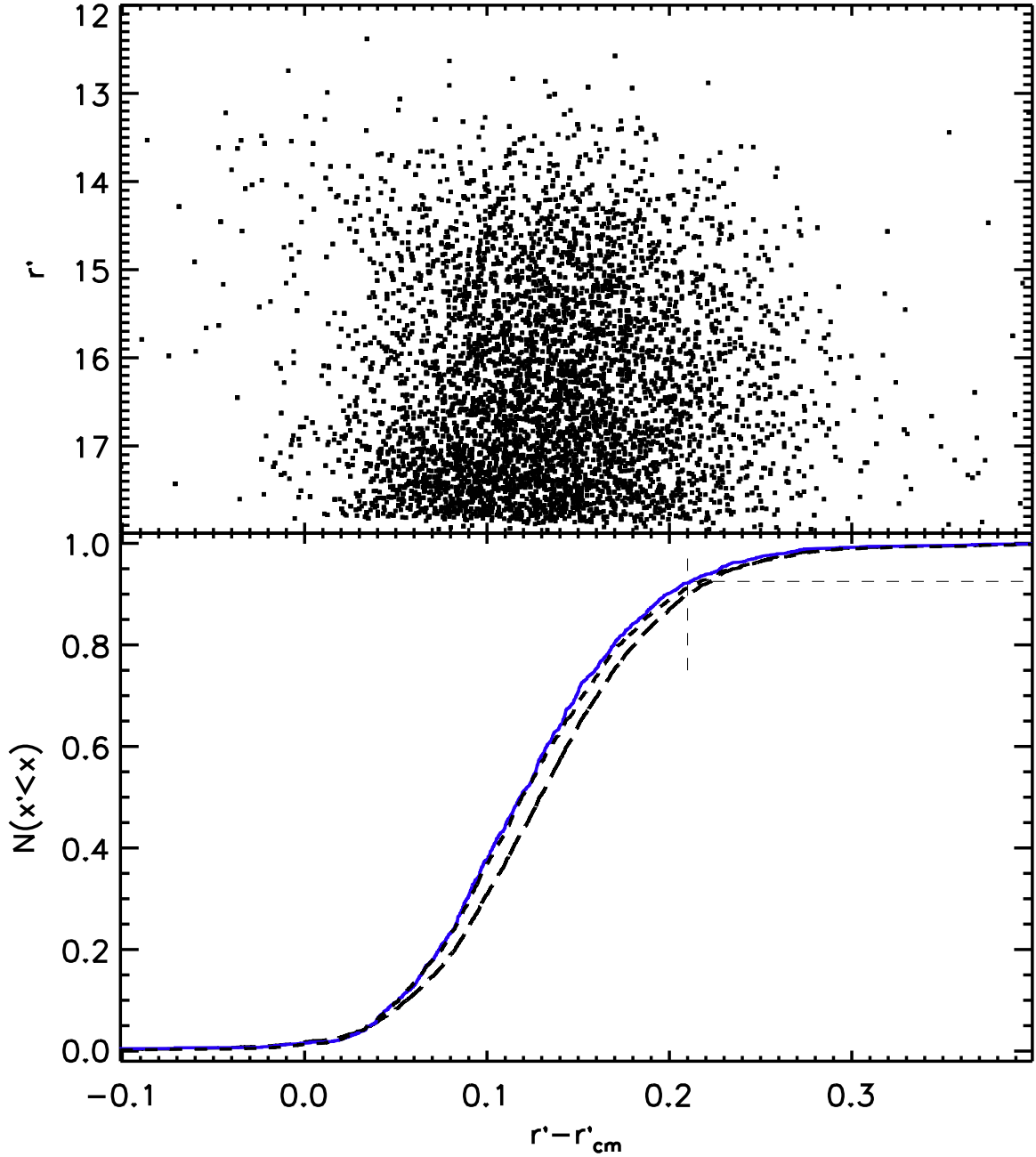}}
\caption{The description of this figure is in the following page.}
\label{r'_rcm_CDF}
\end{figure}

In order to compute $M^{clim}_{r'}$, we assume the completeness limit
of the Main Galaxy Sample $r'$=17.77 and a conservative cut of
$r'-r'_{cm}$=0.21 in the difference between the $r'$-band
composite-model magnitude, $r'_{cm}$, and the $r'$-band petrosian
magnitude $r'$. This cut embraces $\approx$90\% of the overall sample
of cluster galaxies and $\approx$92.5\% of galaxies from the faintest
bin, 17$<$$r'$$<$17.77, or the optical-selected blue galaxies
$u'-r'$$<$2.22, as can be seen in Figure \ref{r'_rcm_CDF}.

\newpage
\addtocounter{figure}{-1}
\begin{figure}[h]
\centering
\caption{{\bf Top panel:} $r'$ vs. $r'-r'_{cm}$ distribution of the
  sample of cluster galaxies. {\bf Bottom panel:} Cumulative
  distribution fraction N(x'$<$x) of $r'-r'_{cm}$ for the sample of
  cluster galaxies. The three curves corresponds to three different
  samples: the long-dashed curve shows the overall sample, the
  continuous blue curve shows the blue galaxies $u'-r'$$<$2.22 and the
  short-dashed curve shows the faintest galaxies
  17$<$$r'$$<$17.77. The vertical and horizontal dashed lines
  correspond, respectively, to a magnitude difference of
  $r'-r'_{cm}$=0.21 and a completeness of 92.5\%.}
\label{r'_rcm_CDF2}
\end{figure}

\newpage

\begin{figure}[p]
\centering
\resizebox{1.00\hsize}{!}{\includegraphics{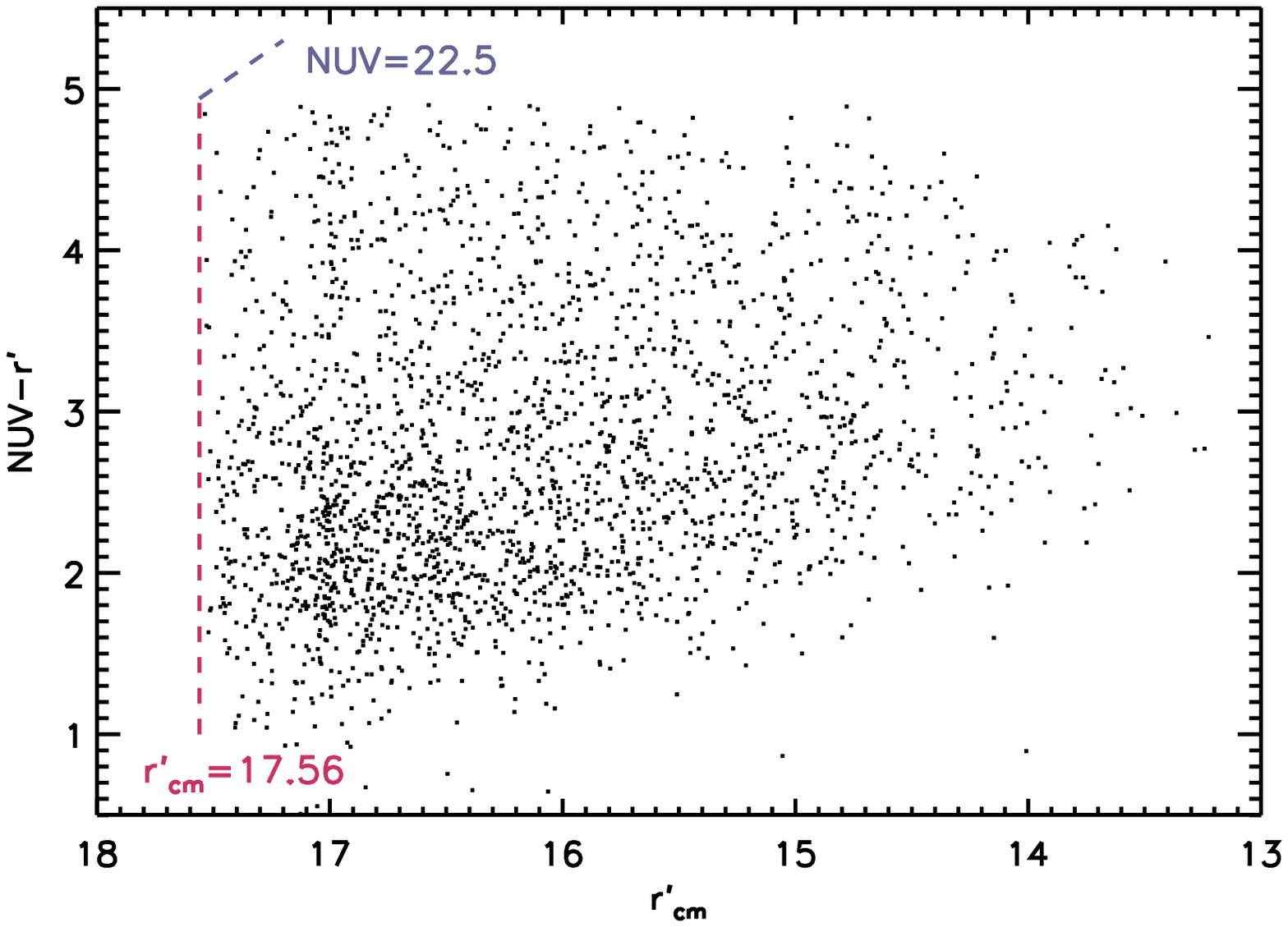}}
\caption{$NUV-r'_{}$ vs. $r'_{cm}$ of the star-forming galaxy
  sample. The red vertical dashed line corresponds to the SDSS MGS
  limit in $r'_{cm}$ magnitude and the blue diagonal dashed line
  corresponds to the GALEX-AIS limit in $NUV$ magnitude.}
\label{NUVr_r}
\end{figure}

Those galaxies for which a $NUV$ detection is not available are
discarded from our analysis. Those galaxies are not observed because
they occupy the gaps between the circular GALEX fields or they are
fainter than the completeness limit of GALEX-AIS. The first case does
not introduce any bias since a correlation between the star formation
properties of those galaxies and their celestial coordinates is not
expected. \textcolor{black}{In the second case, those galaxies are
  outside the color-magnitude locus dedicated to this study as you can
  see in Figure \ref{NUVr_r}, assuming the AB-magnitude limit of a
  SDSS-GALEX matched catalogue $NUV$$\sim$22.5
  \citep{Bianchi_et_al_2007} and the completeness limit of the SDSS
  MGS in the the $r'$-band composite-model magnitude $r'_{cm}$=17.56}


Figure \ref{NUVr_Mr} shows two clearly different behaviors in the
($NUV-r'_{}$)-($M_{r'}$) plane for star-forming galaxies depending on
the environment where they inhabit. The average of $NUV-r'_{}$ for the
field star-forming galaxy sample follows a monotonic trend from
$NUV-r'_{}$$\sim$3.5 at $M_{r'}$$\sim$-22.25 to $NUV-r'_{}$$\sim$2.35
at $M_{r'}$$\sim$-18. In the case of star-forming galaxies inhabiting
the infall regions of clusters, they show a very similar trend to
field star-forming galaxies, with a systematic bias of $\sim$0.1 mag
towards redder values of the $NUV-r'_{}$ average. The trend of
$NUV-r'_{}$ average for star-forming galaxies from the virial regions
show a distinct behavior: their $NUV-r'_{}$ averages in the
$M_{r'}$$<$-20 range are consistent with the $NUV-r'_{}$ trend for
infalling and field star-forming galaxies, within the uncertainties of
$NUV-r'_{}$ averages. In contrast, star-forming galaxies from virial
regions in the -20$<$$M_{r'}$$<$-18 range are systematically biased
toward redder values of $NUV-r'_{}$ in $\sim$0.5 mag.

\begin{figure}[p]
\centering
\resizebox{0.65\vsize}{!}{\includegraphics{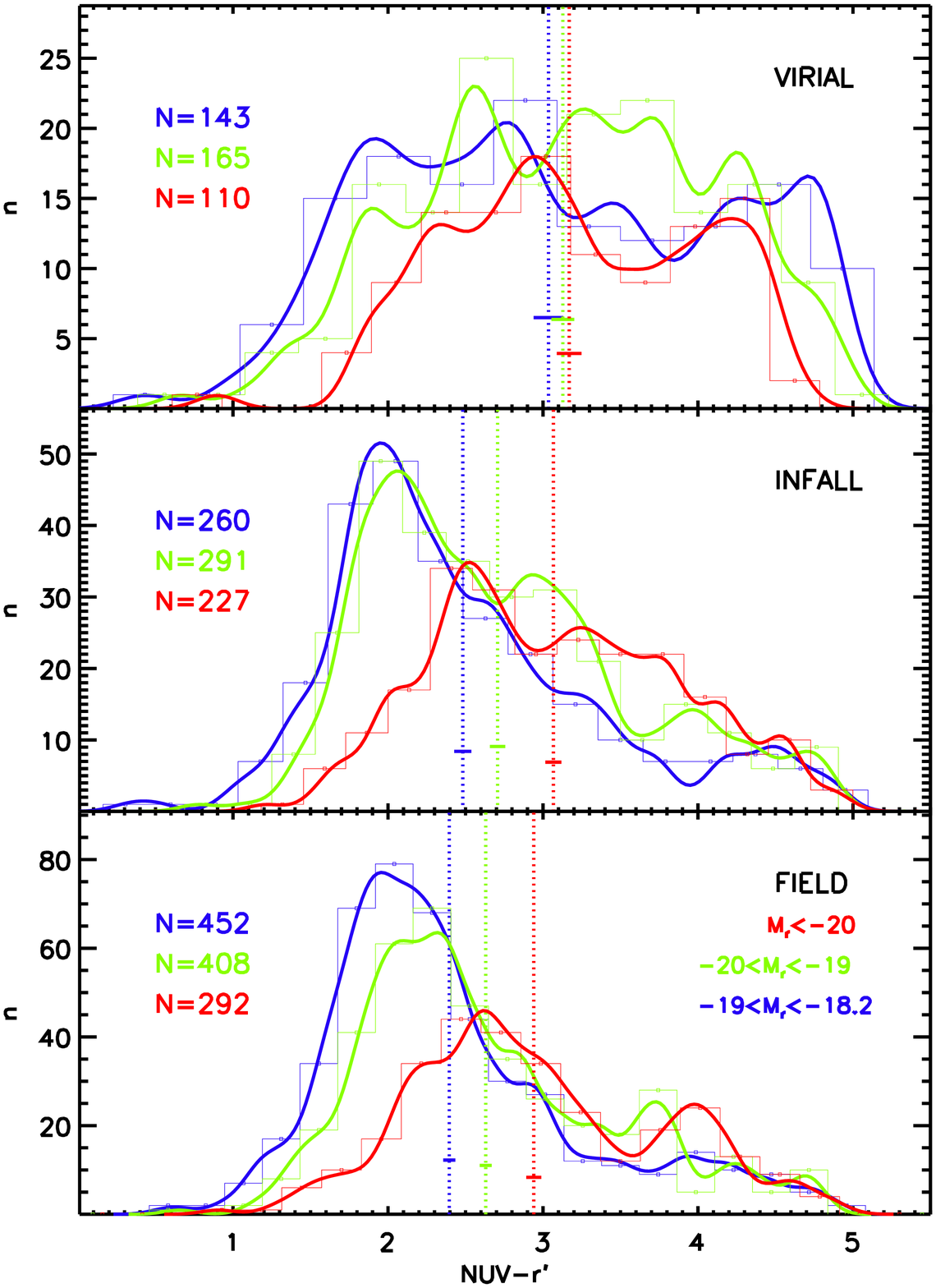}}
\caption{The description of this figure is in the following page.}
\label{NUVr_histo}
\end{figure}

In order to accomplish an analysis in depth of the $NUV-r'_{}$
distribution along the $M_{r'}$ range throughout the defined cluster
regions, we split the star-forming galaxy sample in three luminosity
intervals: the high-luminosity (HL) bin with \mbox{$M_{r'}$$\le$-20}, the
intermediate-luminosity (IL) bin with \mbox{-20$<$$M_{r'}$$\le$-19}, and the
low-luminosity (LL) bin with \mbox{-19$<$$M_{r'}$$\le$-18.2}. We choose the
lowest luminosity cut, $M_{r'}$=-18.2, in order to maximize the size
of the sample of star-forming galaxies in this bin, according to
Eq. \ref{eq:g_comp_rule} and \ref{eq:c_comp_rule}. Figure
\ref{NUVr_histo} shows the $NUV-r'_{}$ distributions of star-forming
galaxies for these three luminosity bins in the three environmental
regions defined above; in Table \ref{med_NUVr} we show the averages
and their bootstrap uncertainties for the distributions of each
subsample.

\newpage
\addtocounter{figure}{-1}
\begin{figure}[h]
\centering
\caption{$NUV-r'_{}$ distributions of star-forming galaxies in each
  luminosity bin in the environmental regions defined above. The three
  different panels correspond to the three different environments,
  from the top to bottom: the virial regions, the cluster infall
  regions and the field environment. Color code identifies the three
  luminosity bins; red \mbox{$M_{r'}$$\le$-20}, green
  \mbox{-20$<$$M_{r'}$$\le$-19} and blue \mbox{-19$<$$M_{r'}$$\le$-18.2}. The thin
  lines correspond to the histograms en each case; the width of the
  bins are computed according to the rule
  $\Delta_{H}$=$\sigma$/$n^{1/5}$ \citep{Turlach_Rev} with $\sigma$
  and $n$ the standard deviation and number of elements of
  distribution, respectively. The solid curves are the kernel density
  estimators computed with a gaussian kernel of
  $\sigma_{gauss}$=$\Delta_{H}$. The vertical dashed lines and the
  horizontal error bar represent, in each case, the average and its
  bootstrap uncertainty for each galaxy subsample,
  respectively. Numbers in the left upper corner of each panel show
  the total number of galaxies in each subsample.}
\label{NUVr_histo2}
\end{figure}

\newpage

As it is shown in Figure \ref{NUVr_histo} and it can be seen in Table
\ref{med_NUVr}, two monotonic trends arise from these results: ({\it
  i}) At all luminosities, the average $NUV-r'_{}$ reddens from the
field environment to the virial regions ({\it ii}) For a given
environment, the average $NUV-r'_{}$ reddens as we move from LL to HL
star-forming galaxies. The first trend, follows the general trend
shown in the color-magnitude diagrams by the "blue cloud"
\citep[e.g.][]{Baldry_et_al_2004,Haines_et_al_2008}. Despite the small
diffferences, the average $NUV-r'_{}$ measured for HL galaxies in the
three environments are consistent within one $\sigma$. This does not
seem to be the case for IL and LL galaxies, where the average
$NUV-r'_{}$ of infall and field galaxies is clearly bluer than that of
virial galaxies, and this behavior is more pronounced for LL than for
IL galaxies. In addition, the $NUV-r'_{}$ distributions for IL and LL
galaxies present a maximum around $NUV-r'_{}$=2 in infall and field
environments. This maximum is clearly absent in the virial
environment, where the $NUV-r'_{}$ distributions present a top-hat
shape at all luminosity ranges.

\begin{table}[p]
\caption{Averages and average uncertainties of $NUV-r'_{}$ distributions
  for each luminosity bin in each environmental region.}
\begin{center}
\begin{tabular}{cccc}
\hline
Environment  & $\langle$$NUV-r'_{}$$\rangle$ & $\pm$ & $\sigma_{bootstrap}$ \\
    (1)      &             (2)           &       &       (3)          \\
\hline
HL \mbox{($M_{r'}$$\le$-20)}   \\
\hline
          
Virial &       3.169 &  $\pm$ &      0.080 \\   
Infall &       3.067 &  $\pm$ &      0.052 \\   
Field  &       2.940 &  $\pm$ &      0.048 \\   
          
\hline
IL \mbox{(-20$<$$M_{r'}$$\le$-19)}   \\
\hline
Virial &       3.128 &  $\pm$  &     0.073 \\  
Infall &       2.707 &  $\pm$  &     0.050 \\  
Field  &       2.630 &  $\pm$  &     0.038 \\  
          
\hline
LL \mbox{(-19$<$$M_{r'}$$\le$-18.2)}   \\
\hline
Virial &       3.035 &  $\pm$  &     0.095 \\  
Infall &       2.483 &  $\pm$  &     0.055 \\  
Field  &       2.395 &  $\pm$  &     0.039 \\  
\hline
\end{tabular}
\end{center}
(1) Environmental region. (2) Mean of $NUV-r'_{}$ distribution in the
corresponding environmental region. (3) Bootstrap uncertainty of the
corresponding $NUV-r'_{}$ mean.
\label{med_NUVr}
\end{table}


\begin{table}[p]
\caption{Results from Kolmogorov-Smirnov test.}
\begin{center}
\begin{tabular}{ccc}
\hline
Subsamples & {\it{D}}$_{n1,n2}$ & {\it{P}}($x_{i1}$,$x_{i2}$) \\
   (1)     &          (2)      &             (3)           \\
\hline
HL \mbox{($M_{r'}$$\le$-20)}           \\
\hline
 Virial - Infall &      0.100 &      0.429  \\ 
 Infall - Field  &      0.117 &     0.0570  \\ 
 Virial - Field  &      0.162 &     0.0256  \\ 
\hline
IL \mbox{(-20$<$$M_{r'}$$\le$-19)}     \\
\hline
 Virial - Infall &      0.245 &   4.75$\cdot$10$^{6}$  \\ 
 Infall - Field  &     0.076 &      0.268            \\ 
 Virial - Field  &      0.295 &   1.56$\cdot$10$^{9}$  \\ 
\hline
LL \mbox{(-19$<$$M_{r'}$$\le$-18.2)} \\
\hline
 Virial - Infall &      0.285 &   3.96$\cdot$10$^{7}$  \\ 
 Infall - Field  &     0.081 &      0.214             \\ 
 Virial - Field  &      0.334 &   3.38$\cdot$10$^{11}$  \\ 
\hline
\end{tabular}
\end{center}
(1) Environmental regions where the K-S is applied to the $NUV-r'_{}$
distributions of their galaxies subsamples. (2) Maximum difference
between the two cumulative distribution functions of the $NUV-r'_{}$
distributions. (3) Probability that the two $NUV-r'_{}$ distributions
came from the same parent population.
\label{K-S_wov}
\end{table}

\begin{figure}[p]
\centering
\resizebox{1.00\hsize}{!}{\includegraphics{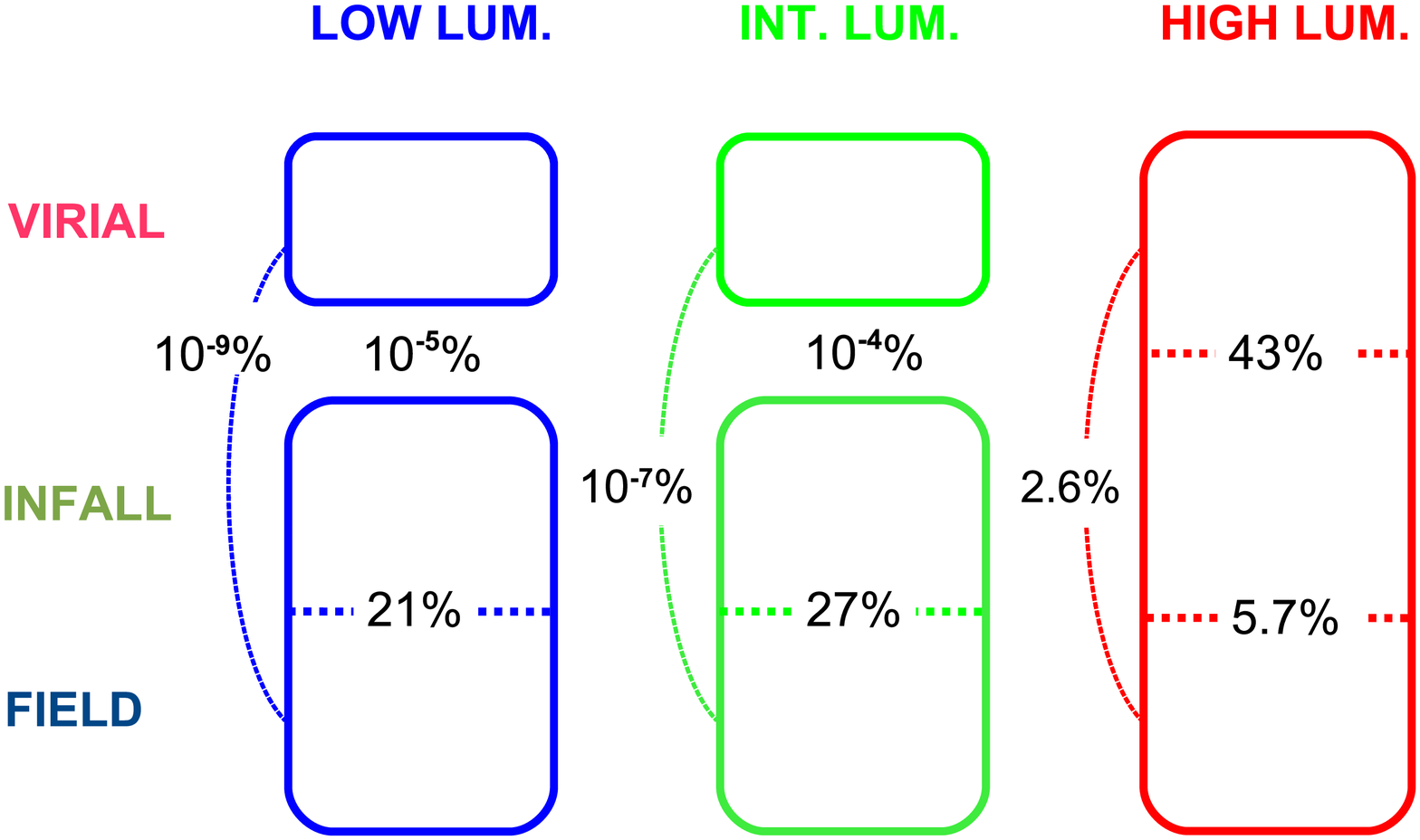}}
\caption{Simplified scheme of the statistical similarity between the
  $NUV-r'_{}$ distributions of star-forming galaxy subsamples of the
  same luminosity bin from the different environmental regions. The
  figures on the picture are the K-S probabilities that the
  distributions came from the same parent population. The star-forming
  galaxy subsamples correspond, from left to right, to the
  low-luminosity bin \mbox{-19$<$$M_{r'}$$\le$-18.2}, the
  intermediate-luminosity bin \mbox{-20$<$$M_{r'}$$\le$-19} and the
  high-luminosity bin \mbox{$M_{r'}$$\le$-20} and; from the top to the
  bottom, to the virial regions, the infall regions and the field
  environment. Monolithic blocks represent of star-forming galaxy
  subsamples presenting a significant similarity in their $NUV-r'_{}$
  distributions (see the text for details).}
\label{KS_graph}
\end{figure}

\section{Discussion}
\label{sec:discussion}

In the previos section we have shown that the $NUV-r'_{}$
distributions of galaxies in different environments present
significant differences depending on the $r'$-band luminosity
considered. In order to accomplish a rigorous analysis based on the
$NUV-r'_{}$ distributions, we apply a Kolmogorov-Smirnov (K-S)
test. Table \ref{K-S_wov} shows the results from the K-S test when
comparing the samples corresponding to the three environments for each
luminosity bin. Figure \ref{KS_graph} illustrates the results of the
K-S test by showing a simplified scheme of the probability that the
star-forming galaxy subsamples of same luminosity bin in different
environments come from the same population. In what follows we discuss
the implications of these results for the different luminosity bins.

\subsection{HL star-forming galaxies}

In the case of HL \mbox{($M_{r'}$$\le$-20)} star-forming galaxies, the
probability $P(x_{1i},x_{2i})$ that the $NUV-r'_{}$ distribution of
the virial regions and the infall regions came from the same parent
population is $P$$\approx$43\%, so they show a high similarity in
their $NUV-r'_{}$ distributions. When comparing the infall regions and
the field environment, this probability lowers to
$P(x_{1i},x_{2i})$$\approx$6\%, so we can not reject the null
hypothesis that the distributions came from the same parent
population. The comparison between the virial regions and the field
environment yields a probability $P(x_{1i},x_{2i})$$\approx$3\%, which
is still low but it does not allow to reject the null hypothesis. All
these results point out that the HL star-forming galaxies, from the
point of view of the environmental influence on their $NUV-r'_{}$
distributions, form a unique family of galaxies only subtle affected
by the environment where they inhabit. They only show a very mild
gradient of reddening towards the inner parts of galaxy
clusters. Assuming that the $NUV-r'_{}$ distributions of the HL
star-forming galaxies as our unique observable, we propose two
scenarios about the environmental influence on the star formation
activity of these galaxies: (1) there is not a strong environmental
influence on the star formation activity of these galaxies in the last
$\sim$10$^{8}$ yr or (2) the environmental processes which would
affect these galaxies act on short time-scales, taking as a reference
the time-scale of UV luminosity, $\tau_{UV}$$\sim$10$^{8}$ yr.

\begin{figure}[p]
\centering
\resizebox{1.00\hsize}{!}{\includegraphics{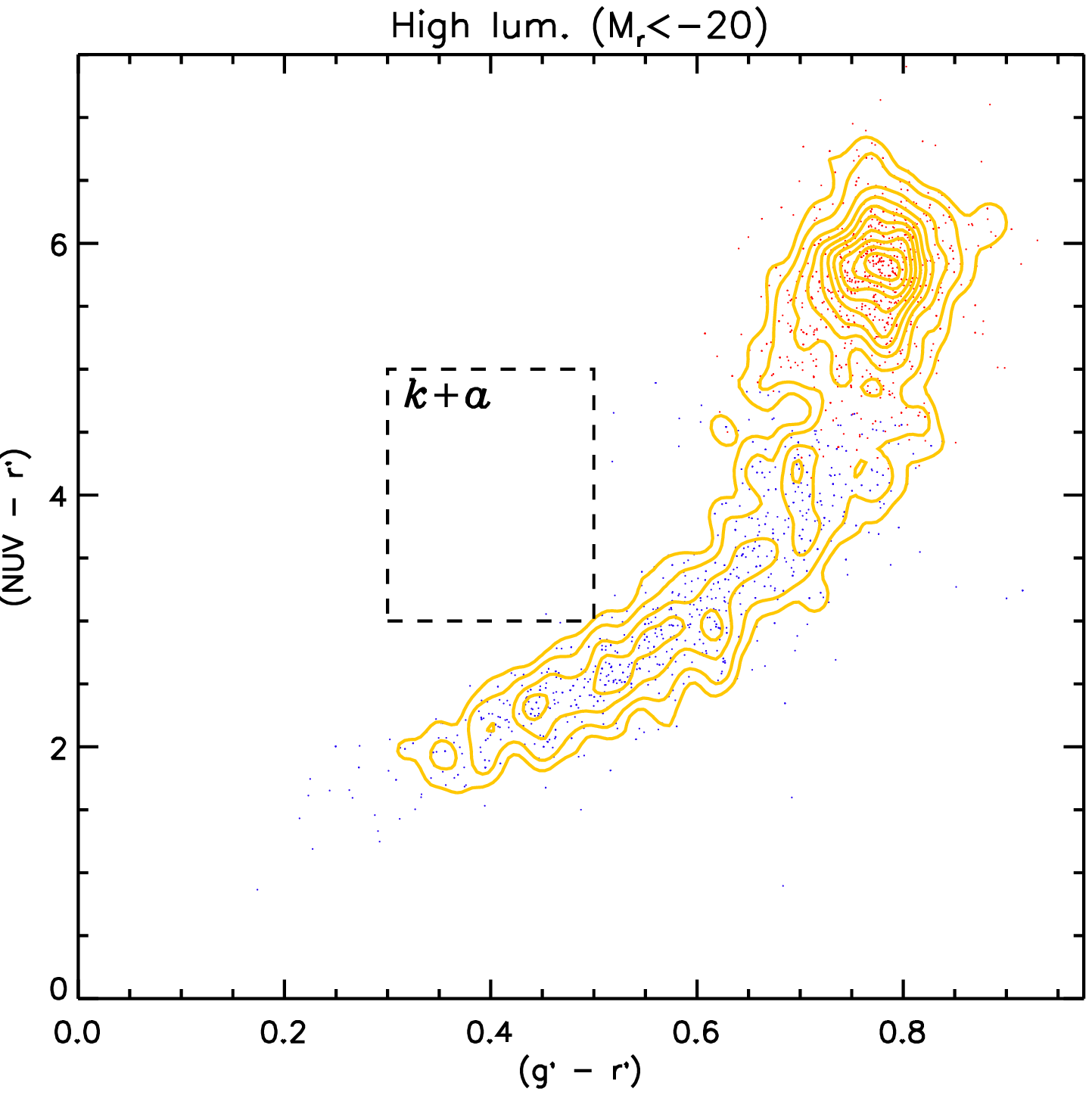}}
\caption{($NUV-r'_{}$) vs. ($g'-r'$) for the $M_{r'}$$<$-20 galaxy
  sample. Yellow isocontours represents the isodensity contours of HL
  galaxies. Blue and red points represent, respectively, star-forming
  and passive galaxies under the prescription shown in Section
  \ref{eq:NUVr_ur_cut}. The square box represents the locus of the
      {\it k+a} galaxies in this color-color diagram
      \citep[Fig. 1]{Kaviraj_E+A}.}
\label{NUVr_gr}
\end{figure}

A sudden truncation of a previously significant star formation in the
last 1.5 Gyr leaves (at least) two observables imprints on the galaxy
spectrum: strong Balmer lines in absorption ($EW_{H\delta}$$>$3 \AA)
and negligible emission lines (e.g. [OII])
\citep{Poggianti_et_al_2004}; this kind of galaxies have been named
      {\it k+a} galaxies \citep[in the modern
        designation,][]{Dressler_et_al_1999} or E+A galaxies \citep[as
        they were first identified by][]{Dressler&Gunn_1983}. Because
      of this, the {\it k+a} spectra are often linked with {\it
        poststarburst} galaxies. Those galaxies have an specific locus
      in an UV-optical color-color diagram (e.g. GALEX-SDSS color
      diagram). Their optical colors are as blue as the blue-cloud
      galaxies and their UV-optical colors are as red as the
      red-sequence galaxies \citep{Kaviraj_E+A}. \textcolor{black}{As
        you can see in Figure \ref{NUVr_gr}, the appearance of our
        sample of HL galaxies in such color-color diagram shows that
        this kind of galaxies are clearly absent in the observed locus
        for {\it k+a} galaxies found by \citet[][see their figure
          1]{Kaviraj_E+A}}. This is in agreement with other results
      found for local galaxy clusters
      \citep[e.g.][]{Fabricant_et_al_1991}. In contrast, the luminous
            {\it k+a} galaxies are abundant in the field
            \citep[e.g.][]{Goto_2005} and also they represent a
            significant fraction of the cluster dwarf galaxy
            population in Coma cluster
            \citep{Poggianti_et_al_2004}. Summarizing, a short
            time-scale mechanism quenching star formation (i.e. gas
            stripping) in HL star-forming galaxies is not expected to
            have affected significantly those galaxies in the last
            $\sim$1.5 Gyr.

\subsection{IL star-forming galaxies}

In the case of IL \mbox{(-20$<$$M_{r'}$$\le$-19)} star-forming galaxies,
there is a high similarity between the $NUV-r'_{}$ distribution of
this subsample of galaxies inhabiting the infall regions and the
corresponding one inhabiting the field environment,
$P(x_{1i},x_{2i})$$\approx$27\%. Contrarily, the comparison of virial
and infall galaxies yields a probability of
$P(x_{1i},x_{2i})$$\sim$10$^{-4}$\%, similar to the one arising from
the comparison of virial and field galaxies
($P(x_{1i},x_{2i})$$\sim$10$^{-7}$\%). These results imply that the
infall and field distributions do not come from the same parent
population as the virial one. These results suggest there is an
environmental mechanism strongly affecting the star formation activity
of the galaxies inside the virial regions on time-scales of 10$^{8}$
yr. Following the scheme proposed by \citet{Treu_et_al_2003} for the
radial range where each environmental process predominantly operates,
we propose starvation through the interaction with the ICM and the DMH
as the main environmental process which affects those galaxies in the
virial regions. In addition, a systematic reddening is observed for
star-forming galaxies in this luminosity bin, \mbox{-20$<$$M_{r'}$$\le$-19},
with respect to the field star-forming galaxies. This is not just
observed in the virial region, but it is observed in the infall region
as well. This result suggests there is some environmental process not
just acting in the virial region but in the infall region too. Again,
following the scheme proposed by \citet{Treu_et_al_2003} for the
radial range for each environmental process, we propose galaxy
harassment as the environmental process quenching the star formation
activity of IL star forming-galaxies in the virial and the infall
regions.

\subsection{LL star-forming galaxies}

The results from the K-S test applied to the LL
\mbox{(-19$<$$M_{r'}$$\le$-18.2)} star-forming galaxies show a similar
scheme to those shown by the IL star-forming galaxies. The LL
star-forming galaxies from the virial regions present a $NUV-r'_{}$
distribution clearly different from those ones from the infall regions
and the field environment. The probability of coming from the same
parent population between virial and infall regions is
$P(x_{1i},x_{2i})$$\sim$10$^{-5}$, while this probability goes down to
$P(x_{1i},x_{2i})$$\sim$10$^{-9}$ for the pair: virial regions
vs. field environment. However, there is a significant similarity
between the $NUV-r'_{}$ distributions from the infall regions and from
the field environment with a probability of
$P(x_{1i},x_{2i})$$\approx$21\%. Even though, these galaxies seem to
suffer a stronger quenching in their star formation activity than the
IL star-forming galaxies. The reddening of the $NUV-r'_{}$ averages
towards the inner parts of the clusters is larger for the LL
star-forming galaxies than for the IL star-forming galaxies, assuming
the galaxies from the virial regions as the reference, including the
gross differences and those normalized to the sum of the
uncertainties, see Table \ref{med_NUVr}. The comparison of the virial
regions with the field environment produces a difference normalized to
the sum of uncertainties $\Delta$/$\sigma$ of 4.449 and 4.767 for the
IL and the LL star-forming galaxies, respectively. The comparison of
the virial regions with the infall regions produces a
$\Delta$/$\sigma$ of 3.42190 and 3.67115 for the IL and for the LL
star-forming galaxies, respectively.

The proposed environmental mechanisms which can affect more strongly
the LL star-forming galaxies seem to be (in the same way as the IL
star-forming galaxies) the galaxy harassment - in order to explain the
reddening in the infall regions with respect to the field environment
- and the starvation, through the interaction with the ICM and through
interaction with the DMH - to explain the evident difference in the
$NUV-r'_{}$ distributions between the virial regions and the other
environmental regions.


As mentioned above, the effects of environment in the quenching of
star formation activity seem to be more pronounced on the LL
star-forming galaxies than on the IL star-forming galaxies. This
suggests that the intensity of the above-cited environmental processes
has some dependence on the galaxy luminosity or on the galaxy (stellar
or gravitational) mass. In the case of galaxy harassment,
\citet{Moore_et_al_1999} found, through equilibrium high-resolution
model of spirals embedded in a N-body simulation, that the response of
a spiral galaxy to tidal encounters depends primarily on the central
depth of its gravitational potential and its disc scale-length, mainly
determined by the stellar component of galaxies. Low-surface
brightness galaxies evolve dramatically under the influence of rapid
encounters with substructures and strong tidal shocks from the global
cluster potential (i.e. galaxy harassment), while high-surface
brightness galaxies are more stable to the galaxy stirring in clusters
and to the tidal encounters. In the case of starvation,
\citet{Bekki_et_al_2002} performed numerical simulations to evaluate
the stripping which the hot halo gas undergoes when subjected to both
the ram pressure of the ICM and the global tidal field of the
cluster. According to this work, the less massive a galaxy is, the
more efficient those mechanisms are (ICM stripping,
\citealt{Bekki_et_al_2002} and tidal halo stripping,
\citealt{Byrd&Valtonen_1990,Henriksen&Byrd_1996,Byrd&Valtonen_2001}). In
conclusion, the proposed environmental processes acting on IL and LL
star-forming galaxies match the fact that they produce the stronger
effects on star formation activity for the LL star-forming galaxies.


\section{Summary and conclusions}
\label{sec:summ&conclu}

In this paper, we expose an original approach which analyzes the
$NUV-r'_{}$ distributions of a large sample of star-forming galaxies
in clusters covering a range of more than four $r'$-band magnitudes
down to \mbox{$M_{r'}$$\sim$-18} and distributed in three distinct
environmental regions: the virial regions, the cluster infall regions
and the field environment. According to the characteristics of each
environmental process, we discuss and propose which environmental
processes affect the star-forming galaxies depending on their
luminosity. In the following summary, we present the main results and
conclusions of this work.

\begin{itemize}

\item The similarity of the $NUV-r'_{}$ distributions of HL
  \mbox{($M_{r'}$$\le$-20)} star-forming galaxies in the different
  environmental regions togehter with the absence of luminous {\it
    k+a} galaxies in our cluster galaxy sample point out that there is
  neihter a strong environmental influence on the star formation
  activity of these galaxies in the last $\sim$10$^{8}$ yr nor a
  sudden truncation of star formation in the last 1.5 Gyr.

\item The $NUV-r'_{}$ distributions of IL
  \mbox{(-20$<$$M_{r'}$$\le$-19)} star-forming galaxies from the
  infall regions and the field environment show a remarkable
  similarity. In contrast, the $NUV-r'_{}$ distributions of IL
  star-forming galaxies in virial regions are clearly distinct from
  the corresponding ones in infall regions and in the field
  environment. This behavior of the $NUV-r'_{}$ distributions of IL
  star-forming galaxies points out the starvation through the combined
  interaction with the ICM and the DMH as the environmental process
  quenching significantly the star formation activity of IL
  star-forming galaxies in the virial regions of clusters. The
  reddening of $NUV-r'_{}$ distributions of these galaxies observed
  not just in the virial regions, but even in the infall regions with
  respect to the field environment is an indication of there is some
  environmental process acting on these galaxies (at least) in the
  infall region. The galaxy harassment as long-time-scale process
  \mbox{$\tau$$\gtrsim$10$^8$} yr is a suitable candidate to be
  responsible of the observed differences in the $NUV-r'_{}$
  distributions.

\item The LL \mbox{(-19$<$$M_{r'}$$\le$-18.2)} star-forming galaxies present
  the same trends of IL star-forming galaxies in their $NUV-r'_{}$
  distributions. Consequently, we propose that the same environmental
  processes in the same environmental regions are affecting the
  star-formation activity of these galaxies. Even more, the
  $NUV-r'_{}$ averages of the LL star-forming galaxies in the virial
  regions show a stronger reddening with respect to the infall regions
  and the field environment than the IL star-forming galaxies. Indeed,
  the observed effect of these environmental processes (starvation,
  harassment) are stronger for low luminosity galaxies.

\end{itemize}

To conclude, we claim that our results highlight the added value of a
sample of cluster galaxies mapping a broad variety of environmental
conditions with a large range in galaxy luminosity and with
multi-wavelength (from UV to FIR) available data.

\section*{Acknowledgments}

J.D.H.F. thanks L'Osservatorio Astronomico di Padova for hospitality
during the stays where this work was started. We want especially thank
Bianca M$^{\b{a}}$ Poggianti for their help and advice during this
stay. J.D.H.F. acknowledges financial support from the Spanish
Ministerio de Ciencia e Innovaci{\'o}n under the FPI grant
BES-2005-7570. We also acknowledge funding by the Spanish PNAYA
project ESTALLIDOS (grants AYA2007-67965-C03-02, AYA2010-21887-C04-01)
and project CSD2006 00070 ``1st Science with GTC'' from the CONSOLIDER
2010 program of the Spanish MICINN.

This publication has made use of the following resources:

\begin{itemize}

\item the NASA/IPAC Extragalactic Database (NED) which is operated by
  the Jet Propulsion Laboratory, California Institute of Technology,
  under contract with the National Aeronautics and Space
  Administration.

\item the Sloan Digital Sky Survey (SDSS) database. Funding for the
  Sloan Digital Sky Survey (SDSS) and SDSS-II has been provided by the
  Alfred P. Sloan Foundation, the Participating Institutions, the
  National Science Foundation, the U.S. Department of Energy, the
  National Aeronautics and Space Administration, the Japanese
  Monbukagakusho, and the Max Planck Society, and the Higher Education
  Funding Council for England. The SDSS Web site is
  http://www.sdss.org/. \\ The SDSS is managed by the Astrophysical
  Research Consortium (ARC) for the Participating Institutions. The
  Participating Institutions are the American Museum of Natural
  History, Astrophysical Institute Potsdam, University of Basel,
  University of Cambridge, Case Western Reserve University, The
  University of Chicago, Drexel University, Fermilab, the Institute
  for Advanced Study, the Japan Participation Group, The Johns Hopkins
  University, the Joint Institute for Nuclear Astrophysics, the Kavli
  Institute for Particle Astrophysics and Cosmology, the Korean
  Scientist Group, the Chinese Academy of Sciences (LAMOST), Los
  Alamos National Laboratory, the Max-Planck-Institute for Astronomy
  (MPIA), the Max-Planck-Institute for Astrophysics (MPA), New Mexico
  State University, Ohio State University, University of Pittsburgh,
  University of Portsmouth, Princeton University, the United States
  Naval Observatory, and the University of Washington.

\item the Galaxy Evolution Explorer (GALEX), which is a NASA mission
  managed by the Jet Propulsion Laboratory and launched in 2003
  April. We gratefully acknowledge NASA's support for the
  construction, operation, and science analysis for the GALEX mission,
  developed in cooperation with the Centre National d'Etudes Spatiales
  of France and the Korean Ministry of Science and Technology.

\end{itemize}

\appendix
\section*{Appendix}
\label{sec:appendix}

In this appendix, we show the results about the same analysis
developed in this work including the central regions of the clusters
ABELL 2197 and WBL 518 as part of the galaxy sample in the infall
regions.

Table \ref{med_NUVr_0} shows the averages (and uncertainties) of
$NUV-r'_{}$ distributions for each luminosity bin of star-forming
galaxies in each environmental region and the table \ref{K-S_wov_0}
show the results of the K-S test applied to these galaxy subsamples.


\begin{table}[p]
\caption{Averages and uncertainties of distributions of $NUV-r'_{}$ environments.}
\begin{center}
\begin{tabular}{cccc}
\hline
Environments & $\langle$$NUV-r'_{}$$\rangle$  & $\pm$ & $\sigma_{bootstrap}$ \\
          
\hline
HL ($M_{r}$$\le$-20)   \\
\hline
          
Virial &       3.169 &  $\pm$ &      0.080 \\ 
Infall &       3.097 &  $\pm$ &      0.049 \\ 
Field  &       2.940 &  $\pm$ &      0.048 \\ 
          
\hline
IL (-20$<$$M_{r}$$\le$-19)   \\
\hline
Virial &       3.128 &  $\pm$ &      0.073 \\ 
Infall &       2.723 &  $\pm$ &      0.051 \\ 
Field  &       2.630 &  $\pm$ &      0.038 \\ 
          
\hline
LL (-19$<$$M_r$$\le$-18.2)   \\
\hline
Virial &       3.035 &  $\pm$ &      0.095 \\ 
Infall &       2.601 &  $\pm$ &      0.052 \\ 
Field  &       2.395 &  $\pm$ &      0.039 \\ 
\hline
\end{tabular}
\end{center}
\label{med_NUVr_0}
\end{table}

As you can see, the data corresponding to infall regions (the only
line modified in this section) shows, systematically, redder
$NUV-r'_{}$ averages than the averages in the previous analysis in
the three luminosity bins. This is in agreement with the motivation of
the extraction of the central regions of these clusters from the
cluster infall regions of their companions; in these excluded regions
inhabit the typical "red" galaxy population from the central cluster
regions. Anyway, the differences are lower than the associated
uncertainties, except for the case of the LL star-forming galaxies
whose difference is $\approx$2$\sigma$.


\begin{table}[p]
\caption{Results from Kolmogorov-Smirnov test, including the central
  regions of ABELL 2197 and WBL 518 as part of the galaxy sample in
  the infall regions.}
\begin{center}
\begin{tabular}{ccc}
\hline
Subsamples & {\it{D}}$_{n1,n2}$ & {\it{P}}($x_{i1}$,$x_{i2}$) \\
   (1)     &          (2)      &             (3)           \\
\hline
HL \mbox{($M_{r'}$$\le$-20)}           \\
\hline
 Virial - Infall &     0.086 &      0.601 \\ 
 Infall - Field  &      0.137 &     0.0106  \\ 
 Virial - Field  &      0.162 &     0.0260  \\ 
\hline
IL \mbox{(-20$<$$M_{r'}$$\le$-19)}     \\
\hline
 Virial - Infall &      0.241 &   5.13$\cdot$10$^{6}$  \\ 
 Infall - Field  &     0.083 &      0.172             \\ 
 Virial - Field  &      0.295 &   1.56$\cdot$10$^{9}$  \\ 
\hline
LL \mbox{(-19$<$$M_{r'}$$\le$-18.2)} \\
\hline
 Virial - Infall &      0.232 &   4.28$\cdot$10$^{5}$  \\ 
 Infall - Field  &      0.130 &    0.00377             \\ 
 Virial - Field  &      0.334 &   3.38$\cdot$10$^{11}$  \\ 
\hline
\end{tabular}
\end{center}
(1) Environmental regions where the K-S is applied to the $NUV-r'_{}$
distributions of their galaxies subsamples. (2) Maximum difference
between the two cumulative distribution functions of the $NUV-r'_{}$
distributions. (3) Probability that the two $NUV-r'_{}$ distributions
came from the same parent population.
\label{K-S_wov_0}
\end{table}

In the case of results from the K-S test, the changes in the
probabilities in each case point out the extraction of clusters ABELL
2197 and WBL 518 from the infall regions of their neighbor clusters as
a good approach. The probabilities of the $NUV-r'_{}$ distributions
of star-forming galaxy subsamples from the virial region and from the
infall regions of came from the same parent population increase with
respect to the previous analysis, while the probabilities of the
$NUV-r'_{}$ distributions of star-forming galaxy subsamples from the
infall regions and the field environment decrease systematically in
the all luminosity bins. These issues suggest the inclusion of ABELL
2197 and WBL 518 in the infall regions makes the $NUV-r'_{}$
distributions more similar between the infall regions and the virial
regions, in the same way it makes the $NUV-r'_{}$ distributions of
the infall regions more different from those from the field
environment. Therefore, the central regions of these clusters have a
higher similarity in the $NUV-r'_{}$ distributions of their
star-forming galaxy population with the virial regions than the infall
regions, as we suppose in our approach.


\bibliographystyle{apj} 
\bibliography{Bibliografia_v2}

\begin{thebibliography}{63}
\expandafter\ifx\csname natexlab\endcsname\relax\def\natexlab#1{#1}\fi

\bibitem[{{Abadi} {et~al.}(1999){Abadi}, {Moore}, \&
  {Bower}}]{Abadi_et_al_1999}
{Abadi}, M.~G., {Moore}, B., \& {Bower}, R.~G. 1999, \mnras, 308, 947

\bibitem[{{Abazajian} {et~al.}(2004){Abazajian}, {Adelman-McCarthy},
  {Ag{\"u}eros}, {Allam}, {Anderson}, {Anderson}, {Annis}, {Bahcall}, {Baldry},
  {Bastian}, {Berlind}, {Bernardi}, {Blanton}, {Bochanski}, {Boroski},
  {Briggs}, {Brinkmann}, {Brunner}, {Budav{\'a}ri}, {Carey}, {Carliles},
  {Castander}, {Connolly}, {Csabai}, {Doi}, {Dong}, {Eisenstein}, {Evans},
  {Fan}, {Finkbeiner}, {Friedman}, {Frieman}, {Fukugita}, {Gal}, {Gillespie},
  {Glazebrook}, {Gray}, {Grebel}, {Gunn}, {Gurbani}, {Hall}, {Hamabe},
  {Harris}, {Harris}, {Harvanek}, {Heckman}, {Hendry}, {Hennessy}, {Hindsley},
  {Hogan}, {Hogg}, {Holmgren}, {Ichikawa}, {Ichikawa}, {Ivezi{\'c}}, {Jester},
  {Johnston}, {Jorgensen}, {Kent}, {Kleinman}, {Knapp}, {Kniazev}, {Kron},
  {Krzesinski}, {Kunszt}, {Kuropatkin}, {Lamb}, {Lampeitl}, {Lee}, {Leger},
  {Li}, {Lin}, {Loh}, {Long}, {Loveday}, {Lupton}, {Malik}, {Margon},
  {Matsubara}, {McGehee}, {McKay}, {Meiksin}, {Munn}, {Nakajima}, {Nash},
  {Neilsen}, {Newberg}, {Newman}, {Nichol}, {Nicinski}, {Nieto-Santisteban},
  {Nitta}, {Okamura}, {O'Mullane}, {Ostriker}, {Owen}, {Padmanabhan},
  {Peoples}, {Pier}, {Pope}, {Quinn}, {Richards}, {Richmond}, {Rix}, {Rockosi},
  {Schlegel}, {Schneider}, {Scranton}, {Sekiguchi}, {Seljak}, {Sergey},
  {Sesar}, {Sheldon}, {Shimasaku}, {Siegmund}, {Silvestri}, {Smith}, {Smol{\v
  c}i{\'c}}, {Snedden}, {Stebbins}, {Stoughton}, {Strauss}, {SubbaRao},
  {Szalay}, {Szapudi}, {Szkody}, {Szokoly}, {Tegmark}, {Teodoro}, {Thakar},
  {Tremonti}, {Tucker}, {Uomoto}, {Vanden Berk}, {Vandenberg}, {Vogeley},
  {Voges}, {Vogt}, {Walkowicz}, {Wang}, {Weinberg}, {West}, {White}, {Wilhite},
  {Xu}, {Yanny}, {Yasuda}, {Yip}, {Yocum}, {York}, {Zehavi}, {Zibetti}, \&
  {Zucker}}]{SDSS_II}
{Abazajian}, K., {Adelman-McCarthy}, J.~K., {Ag{\"u}eros}, M.~A., {et~al.}
  2004, \aj, 128, 502

\bibitem[{{Adelman-McCarthy} {et~al.}(2008){Adelman-McCarthy}, {Ag{\"u}eros},
  {Allam}, {Allende Prieto}, {Anderson}, {Anderson}, {Annis}, {Bahcall},
  {Bailer-Jones}, {Baldry}, {Barentine}, {Bassett}, {Becker}, {Beers}, {Bell},
  {Berlind}, {Bernardi}, {Blanton}, {Bochanski}, {Boroski}, {Brinchmann},
  {Brinkmann}, {Brunner}, {Budav{\'a}ri}, {Carliles}, {Carr}, {Castander},
  {Cinabro}, {Cool}, {Covey}, {Csabai}, {Cunha}, {Davenport}, {Dilday}, {Doi},
  {Eisenstein}, {Evans}, {Fan}, {Finkbeiner}, {Friedman}, {Frieman},
  {Fukugita}, {G{\"a}nsicke}, {Gates}, {Gillespie}, {Glazebrook}, {Gray},
  {Grebel}, {Gunn}, {Gurbani}, {Hall}, {Harding}, {Harvanek}, {Hawley},
  {Hayes}, {Heckman}, {Hendry}, {Hindsley}, {Hirata}, {Hogan}, {Hogg}, {Hyde},
  {Ichikawa}, {Ivezi{\'c}}, {Jester}, {Johnson}, {Jorgensen}, {Juri{\'c}},
  {Kent}, {Kessler}, {Kleinman}, {Knapp}, {Kron}, {Krzesinski}, {Kuropatkin},
  {Lamb}, {Lampeitl}, {Lebedeva}, {Lee}, {Leger}, {L{\'e}pine}, {Lima}, {Lin},
  {Long}, {Loomis}, {Loveday}, {Lupton}, {Malanushenko}, {Malanushenko},
  {Mandelbaum}, {Margon}, {Marriner}, {Mart{\'{\i}}nez-Delgado}, {Matsubara},
  {McGehee}, {McKay}, {Meiksin}, {Morrison}, {Munn}, {Nakajima}, {Neilsen},
  {Newberg}, {Nichol}, {Nicinski}, {Nieto-Santisteban}, {Nitta}, {Okamura},
  {Owen}, {Oyaizu}, {Padmanabhan}, {Pan}, {Park}, {Peoples}, {Pier}, {Pope},
  {Purger}, {Raddick}, {Re Fiorentin}, {Richards}, {Richmond}, {Riess}, {Rix},
  {Rockosi}, {Sako}, {Schlegel}, {Schneider}, {Schreiber}, {Schwope}, {Seljak},
  {Sesar}, {Sheldon}, {Shimasaku}, {Sivarani}, {Smith}, {Snedden}, {Steinmetz},
  {Strauss}, {SubbaRao}, {Suto}, {Szalay}, {Szapudi}, {Szkody}, {Tegmark},
  {Thakar}, {Tremonti}, {Tucker}, {Uomoto}, {Vanden Berk}, {Vandenberg},
  {Vidrih}, {Vogeley}, {Voges}, {Vogt}, {Wadadekar}, {Weinberg}, {West},
  {White}, {Wilhite}, {Yanny}, {Yocum}, {York}, {Zehavi}, \&
  {Zucker}}]{Adelman-McCarthy_et_al_2008}
{Adelman-McCarthy}, J.~K., {Ag{\"u}eros}, M.~A., {Allam}, S.~S., {et~al.} 2008,
  \apjs, 175, 297

\bibitem[{{Baldry} {et~al.}(2004){Baldry}, {Glazebrook}, {Brinkmann},
  {Ivezi{\'c}}, {Lupton}, {Nichol}, \& {Szalay}}]{Baldry_et_al_2004}
{Baldry}, I.~K., {Glazebrook}, K., {Brinkmann}, J., {et~al.} 2004, \apj, 600,
  681

\bibitem[{{Bekki}(1998)}]{Bekki_1998}
{Bekki}, K. 1998, \apjl, 502, L133

\bibitem[{{Bekki} {et~al.}(2002){Bekki}, {Couch}, \&
  {Shioya}}]{Bekki_et_al_2002}
{Bekki}, K., {Couch}, W.~J., \& {Shioya}, Y. 2002, \apj, 577, 651

\bibitem[{{Bertin} \& {Arnouts}(1996)}]{Bertin&Arnouts_1996}
{Bertin}, E. \& {Arnouts}, S. 1996, \aaps, 117, 393

\bibitem[{{Bianchi} {et~al.}(2007){Bianchi}, {Rodriguez-Merino}, {Viton},
  {Laget}, {Efremova}, {Herald}, {Conti}, {Shiao}, {Gil de Paz}, {Salim},
  {Thakar}, {Friedman}, {Rey}, {Thilker}, {Barlow}, {Budav{\'a}ri}, {Donas},
  {Forster}, {Heckman}, {Lee}, {Madore}, {Martin}, {Milliard}, {Morrissey},
  {Neff}, {Rich}, {Schiminovich}, {Seibert}, {Small}, {Szalay}, {Wyder},
  {Welsh}, \& {Yi}}]{Bianchi_et_al_2007}
{Bianchi}, L., {Rodriguez-Merino}, L., {Viton}, M., {et~al.} 2007, \apjs, 173,
  659

\bibitem[{{Boselli} \& {Gavazzi}(2006)}]{Boselli&Gavazzi_2006}
{Boselli}, A. \& {Gavazzi}, G. 2006, \pasp, 118, 517

\bibitem[{{Buat} {et~al.}(2005){Buat}, {Iglesias-P{\'a}ramo}, {Seibert},
  {Burgarella}, {Charlot}, {Martin}, {Xu}, {Heckman}, {Boissier}, {Boselli},
  {Barlow}, {Bianchi}, {Byun}, {Donas}, {Forster}, {Friedman}, {Jelinski},
  {Lee}, {Madore}, {Malina}, {Milliard}, {Morissey}, {Neff}, {Rich},
  {Schiminovitch}, {Siegmund}, {Small}, {Szalay}, {Welsh}, \&
  {Wyder}}]{Buat_et_al_2005}
{Buat}, V., {Iglesias-P{\'a}ramo}, J., {Seibert}, M., {et~al.} 2005, \apjl,
  619, L51

\bibitem[{{Byrd} \& {Valtonen}(1990)}]{Byrd&Valtonen_1990}
{Byrd}, G. \& {Valtonen}, M. 1990, \apj, 350, 89

\bibitem[{{Byrd} \& {Valtonen}(2001)}]{Byrd&Valtonen_2001}
{Byrd}, G. \& {Valtonen}, M. 2001, \aj, 121, 2943

\bibitem[{{Chilingarian} \& {Zolotukhin}(2012)}]{Chilingarian&Zolotukhin_2011}
{Chilingarian}, I.~V. \& {Zolotukhin}, I.~Y. 2012, \mnras, 419, 1727

\bibitem[{{Cortese} {et~al.}(2004){Cortese}, {Gavazzi}, {Boselli},
  {Iglesias-Paramo}, \& {Carrasco}}]{Cortese_et_al_2004}
{Cortese}, L., {Gavazzi}, G., {Boselli}, A., {Iglesias-Paramo}, J., \&
  {Carrasco}, L. 2004, \aap, 425, 429

\bibitem[{{den Hartog} \& {Katgert}(1996)}]{den_Hartog&Katgert_1996}
{den Hartog}, R. \& {Katgert}, P. 1996, \mnras, 279, 349

\bibitem[{{Diaferio}(1999)}]{Diaferio_1999}
{Diaferio}, A. 1999, \mnras, 309, 610

\bibitem[{{Dressler}(1980)}]{Dressler_1980}
{Dressler}, A. 1980, \apj, 236, 351

\bibitem[{{Dressler} \& {Gunn}(1983)}]{Dressler&Gunn_1983}
{Dressler}, A. \& {Gunn}, J.~E. 1983, \apj, 270, 7

\bibitem[{{Dressler} {et~al.}(1999){Dressler}, {Smail}, {Poggianti}, {Butcher},
  {Couch}, {Ellis}, \& {Oemler}}]{Dressler_et_al_1999}
{Dressler}, A., {Smail}, I., {Poggianti}, B.~M., {et~al.} 1999, \apjs, 122, 51

\bibitem[{{Evrard}(1991)}]{Evrard_1991}
{Evrard}, A.~E. 1991, \mnras, 248, 8P

\bibitem[{{Fabricant} {et~al.}(1991){Fabricant}, {McClintock}, \&
  {Bautz}}]{Fabricant_et_al_1991}
{Fabricant}, D.~G., {McClintock}, J.~E., \& {Bautz}, M.~W. 1991, \apj, 381, 33

\bibitem[{{Fujita}(1998)}]{Fujita_1998}
{Fujita}, Y. 1998, \apj, 509, 587

\bibitem[{{Ghigna} {et~al.}(1998){Ghigna}, {Moore}, {Governato}, {Lake},
  {Quinn}, \& {Stadel}}]{Ghigna_et_al_1998}
{Ghigna}, S., {Moore}, B., {Governato}, F., {et~al.} 1998, \mnras, 300, 146

\bibitem[{{G{\'o}mez} {et~al.}(2003){G{\'o}mez}, {Nichol}, {Miller}, {Balogh},
  {Goto}, {Zabludoff}, {Romer}, {Bernardi}, {Sheth}, {Hopkins}, {Castander},
  {Connolly}, {Schneider}, {Brinkmann}, {Lamb}, {SubbaRao}, \&
  {York}}]{Gomez_et_al_2003}
{G{\'o}mez}, P.~L., {Nichol}, R.~C., {Miller}, C.~J., {et~al.} 2003, \apj, 584,
  210

\bibitem[{{Goto}(2005)}]{Goto_2005}
{Goto}, T. 2005, \mnras, 357, 937

\bibitem[{{Haines} {et~al.}(2008){Haines}, {Gargiulo}, \&
  {Merluzzi}}]{Haines_et_al_2008}
{Haines}, C.~P., {Gargiulo}, A., \& {Merluzzi}, P. 2008, \mnras, 385, 1201

\bibitem[{{Henriksen} \& {Byrd}(1996)}]{Henriksen&Byrd_1996}
{Henriksen}, M. \& {Byrd}, G. 1996, \apj, 459, 82

\bibitem[{{{Hern{\'a}ndez-Fern{\'a}ndez} et
  al.}(2012)}]{Hernandez-Fernandez_et_al_2011_GC}
{{Hern{\'a}ndez-Fern{\'a}ndez} et al.} 2012, {{UV to FIR catalogue of a galaxy
  sample in nearby clusters: SEDs and environmental trends} (accepted,
  arXiv:1201.2697)}

\bibitem[{{Icke}(1985)}]{Icke_1985}
{Icke}, V. 1985, \aap, 144, 115

\bibitem[{{Iglesias-P{\'a}ramo} {et~al.}(2006){Iglesias-P{\'a}ramo}, {Buat},
  {Takeuchi}, {Xu}, {Boissier}, {Boselli}, {Burgarella}, {Madore}, {Gil de
  Paz}, {Bianchi}, {Barlow}, {Byun}, {Donas}, {Forster}, {Friedman}, {Heckman},
  {Jelinski}, {Lee}, {Malina}, {Martin}, {Milliard}, {Morrissey}, {Neff},
  {Rich}, {Schiminovich}, {Seibert}, {Siegmund}, {Small}, {Szalay}, {Welsh}, \&
  {Wyder}}]{Iglesias-Paramo_et_al_2006}
{Iglesias-P{\'a}ramo}, J., {Buat}, V., {Takeuchi}, T.~T., {et~al.} 2006, \apjs,
  164, 38

\bibitem[{{Kaiser}(1987)}]{Kaiser_1987}
{Kaiser}, N. 1987, \mnras, 227, 1

\bibitem[{{Kauffmann} {et~al.}(2007){Kauffmann}, {Heckman}, {Budav{\'a}ri},
  {Charlot}, {Hoopes}, {Martin}, {Seibert}, {Barlow}, {Bianchi}, {Conrow},
  {Donas}, {Forster}, {Friedman}, {Lee}, {Madore}, {Milliard}, {Morrissey},
  {Neff}, {Rich}, {Schiminovich}, {Small}, {Szalay}, {Wyder}, \&
  {Yi}}]{Kauffmann_et_al_2007}
{Kauffmann}, G., {Heckman}, T.~M., {Budav{\'a}ri}, T., {et~al.} 2007, \apjs,
  173, 357

\bibitem[{{Kaviraj} {et~al.}(2007{\natexlab{a}}){Kaviraj}, {Kirkby}, {Silk}, \&
  {Sarzi}}]{Kaviraj_E+A}
{Kaviraj}, S., {Kirkby}, L.~A., {Silk}, J., \& {Sarzi}, M. 2007{\natexlab{a}},
  \mnras, 382, 960

\bibitem[{{Kaviraj} {et~al.}(2007{\natexlab{b}}){Kaviraj}, {Rey}, {Rich},
  {Yoon}, \& {Yi}}]{Kaviraj_et_al_2007}
{Kaviraj}, S., {Rey}, S., {Rich}, R.~M., {Yoon}, S., \& {Yi}, S.~K.
  2007{\natexlab{b}}, \mnras, 381, L74

\bibitem[{{Kennicutt}(1998)}]{Kennicutt_1998}
{Kennicutt}, Jr., R.~C. 1998, \araa, 36, 189

\bibitem[{{Larson} {et~al.}(1980){Larson}, {Tinsley}, \&
  {Caldwell}}]{Larson_et_al_1980}
{Larson}, R.~B., {Tinsley}, B.~M., \& {Caldwell}, C.~N. 1980, \apj, 237, 692

\bibitem[{{Leitherer} {et~al.}(1999){Leitherer}, {Schaerer}, {Goldader},
  {Gonz{\'a}lez Delgado}, {Robert}, {Kune}, {de Mello}, {Devost}, \&
  {Heckman}}]{Leitherer_et_al_1999}
{Leitherer}, C., {Schaerer}, D., {Goldader}, J.~D., {et~al.} 1999, \apjs, 123,
  3

\bibitem[{{Lewis} {et~al.}(2002){Lewis}, {Balogh}, {De Propris}, {Couch},
  {Bower}, {Offer}, {Bland-Hawthorn}, {Baldry}, {Baugh}, {Bridges}, {Cannon},
  {Cole}, {Colless}, {Collins}, {Cross}, {Dalton}, {Driver}, {Efstathiou},
  {Ellis}, {Frenk}, {Glazebrook}, {Hawkins}, {Jackson}, {Lahav}, {Lumsden},
  {Maddox}, {Madgwick}, {Norberg}, {Peacock}, {Percival}, {Peterson},
  {Sutherland}, \& {Taylor}}]{Lewis_et_al_2002}
{Lewis}, I., {Balogh}, M., {De Propris}, R., {et~al.} 2002, \mnras, 334, 673

\bibitem[{{Mahdavi} {et~al.}(2000){Mahdavi}, {B{\"o}hringer}, {Geller}, \&
  {Ramella}}]{Mahdavi_et_al_2000}
{Mahdavi}, A., {B{\"o}hringer}, H., {Geller}, M.~J., \& {Ramella}, M. 2000,
  \apj, 534, 114

\bibitem[{{Mahdavi} \& {Geller}(2001)}]{Mahdavi&Geller_2001}
{Mahdavi}, A. \& {Geller}, M.~J. 2001, \apjl, 554, L129

\bibitem[{{Martin} {et~al.}(2005){Martin}, {Fanson}, {Schiminovich},
  {Morrissey}, {Friedman}, {Barlow}, {Conrow}, {Grange}, {Jelinsky},
  {Milliard}, {Siegmund}, {Bianchi}, {Byun}, {Donas}, {Forster}, {Heckman},
  {Lee}, {Madore}, {Malina}, {Neff}, {Rich}, {Small}, {Surber}, {Szalay},
  {Welsh}, \& {Wyder}}]{Martin_et_al_2005}
{Martin}, D.~C., {Fanson}, J., {Schiminovich}, D., {et~al.} 2005, \apjl, 619,
  L1

\bibitem[{{Merritt}(1984)}]{Merritt_1984}
{Merritt}, D. 1984, \apj, 276, 26

\bibitem[{{Mihos}(1995)}]{Mihos_1995}
{Mihos}, J.~C. 1995, \apjl, 438, L75

\bibitem[{{Mihos}(2004)}]{Mihos_2004}
{Mihos}, J.~C. 2004, Clusters of Galaxies: Probes of Cosmological Structure and
  Galaxy Evolution, 277

\bibitem[{{Mihos} \& {Hernquist}(1994)}]{Mihos&Hernquist_1994_ULSB}
{Mihos}, J.~C. \& {Hernquist}, L. 1994, \apjl, 431, L9

\bibitem[{{Mihos} \& {Hernquist}(1996)}]{Mihos&Hernquist_1996}
{Mihos}, J.~C. \& {Hernquist}, L. 1996, \apj, 464, 641

\bibitem[{{Mo} {et~al.}(2010){Mo}, {van den Bosch}, \& {White}}]{Mo_et_al_2010}
{Mo}, H., {van den Bosch}, F.~C., \& {White}, S. 2010, {Galaxy Formation and
  Evolution} (Cambridge University Press. New York)

\bibitem[{{Moore} {et~al.}(1996){Moore}, {Katz}, {Lake}, {Dressler}, \&
  {Oemler}}]{Moore_et_al_1996}
{Moore}, B., {Katz}, N., {Lake}, G., {Dressler}, A., \& {Oemler}, A. 1996,
  \nat, 379, 613

\bibitem[{{Moore} {et~al.}(1998){Moore}, {Lake}, \& {Katz}}]{Moore_et_al_1998}
{Moore}, B., {Lake}, G., \& {Katz}, N. 1998, \apj, 495, 139

\bibitem[{{Moore} {et~al.}(1999){Moore}, {Lake}, {Quinn}, \&
  {Stadel}}]{Moore_et_al_1999}
{Moore}, B., {Lake}, G., {Quinn}, T., \& {Stadel}, J. 1999, \mnras, 304, 465

\bibitem[{{Natarajan} {et~al.}(1998){Natarajan}, {Kneib}, {Smail}, \&
  {Ellis}}]{Natarajan_et_al_1998}
{Natarajan}, P., {Kneib}, J., {Smail}, I., \& {Ellis}, R.~S. 1998, \apj, 499,
  600

\bibitem[{{Poggianti}(2006)}]{Poggianti_2006}
{Poggianti}, B.~M. 2006, in The Many Scales in the Universe: JENAM 2004
  Astrophysics Reviews, ed. {J.~C.~Del Toro Iniesta, E.~J.~Alfaro,
  J.~G.~Gorgas, E.~Salvador-Sole, \& H.~Butcher}, 71

\bibitem[{{Poggianti} {et~al.}(2004){Poggianti}, {Bridges}, {Komiyama}, {Yagi},
  {Carter}, {Mobasher}, {Okamura}, \& {Kashikawa}}]{Poggianti_et_al_2004}
{Poggianti}, B.~M., {Bridges}, T.~J., {Komiyama}, Y., {et~al.} 2004, \apj, 601,
  197

\bibitem[{{Quilis} {et~al.}(2000){Quilis}, {Moore}, \&
  {Bower}}]{Quilis_et_al_2000}
{Quilis}, V., {Moore}, B., \& {Bower}, R. 2000, Science, 288, 1617

\bibitem[{{Rines} {et~al.}(2003){Rines}, {Geller}, {Kurtz}, \&
  {Diaferio}}]{CAIRNS_I}
{Rines}, K., {Geller}, M.~J., {Kurtz}, M.~J., \& {Diaferio}, A. 2003, \aj, 126,
  2152

\bibitem[{{Rines} {et~al.}(2005){Rines}, {Geller}, {Kurtz}, \&
  {Diaferio}}]{Rines_et_al_2005}
{Rines}, K., {Geller}, M.~J., {Kurtz}, M.~J., \& {Diaferio}, A. 2005, \aj, 130,
  1482

\bibitem[{{Salim} {et~al.}(2005){Salim}, {Charlot}, {Rich}, {Kauffmann},
  {Heckman}, {Barlow}, {Bianchi}, {Byun}, {Donas}, {Forster}, {Friedman},
  {Jelinsky}, {Lee}, {Madore}, {Malina}, {Martin}, {Milliard}, {Morrissey},
  {Neff}, {Schiminovich}, {Seibert}, {Siegmund}, {Small}, {Szalay}, {Welsh}, \&
  {Wyder}}]{Salim_et_al_2005}
{Salim}, S., {Charlot}, S., {Rich}, R.~M., {et~al.} 2005, \apjl, 619, L39

\bibitem[{{Stoughton} {et~al.}(2002){Stoughton}, {Lupton}, {Bernardi},
  {Blanton}, {Burles}, {Castander}, {Connolly}, {Eisenstein}, {Frieman},
  {Hennessy}, {Hindsley}, {Ivezi{\'c}}, {Kent}, {Kunszt}, {Lee}, {Meiksin},
  {Munn}, {Newberg}, {Nichol}, {Nicinski}, {Pier}, {Richards}, {Richmond},
  {Schlegel}, {Smith}, {Strauss}, {SubbaRao}, {Szalay}, {Thakar}, {Tucker},
  {Vanden Berk}, {Yanny}, {Adelman}, {Anderson}, {Anderson}, {Annis},
  {Bahcall}, {Bakken}, {Bartelmann}, {Bastian}, {Bauer}, {Berman},
  {B{\"o}hringer}, {Boroski}, {Bracker}, {Briegel}, {Briggs}, {Brinkmann},
  {Brunner}, {Carey}, {Carr}, {Chen}, {Christian}, {Colestock}, {Crocker},
  {Csabai}, {Czarapata}, {Dalcanton}, {Davidsen}, {Davis}, {Dehnen},
  {Dodelson}, {Doi}, {Dombeck}, {Donahue}, {Ellman}, {Elms}, {Evans}, {Eyer},
  {Fan}, {Federwitz}, {Friedman}, {Fukugita}, {Gal}, {Gillespie}, {Glazebrook},
  {Gray}, {Grebel}, {Greenawalt}, {Greene}, {Gunn}, {de Haas}, {Haiman},
  {Haldeman}, {Hall}, {Hamabe}, {Hansen}, {Harris}, {Harris}, {Harvanek},
  {Hawley}, {Hayes}, {Heckman}, {Helmi}, {Henden}, {Hogan}, {Hogg}, {Holmgren},
  {Holtzman}, {Huang}, {Hull}, {Ichikawa}, {Ichikawa}, {Johnston}, {Kauffmann},
  {Kim}, {Kimball}, {Kinney}, {Klaene}, {Kleinman}, {Klypin}, {Knapp},
  {Korienek}, {Krolik}, {Kron}, {Krzesi{\'n}ski}, {Lamb}, {Leger},
  {Limmongkol}, {Lindenmeyer}, {Long}, {Loomis}, {Loveday}, {MacKinnon},
  {Mannery}, {Mantsch}, {Margon}, {McGehee}, {McKay}, {McLean}, {Menou},
  {Merelli}, {Mo}, {Monet}, {Nakamura}, {Narayanan}, {Nash}, {Neilsen},
  {Newman}, {Nitta}, {Odenkirchen}, {Okada}, {Okamura}, {Ostriker}, {Owen},
  {Pauls}, {Peoples}, {Peterson}, {Petravick}, {Pope}, {Pordes}, {Postman},
  {Prosapio}, {Quinn}, {Rechenmacher}, {Rivetta}, {Rix}, {Rockosi}, {Rosner},
  {Ruthmansdorfer}, {Sandford}, {Schneider}, {Scranton}, {Sekiguchi}, {Sergey},
  {Sheth}, {Shimasaku}, {Smee}, {Snedden}, {Stebbins}, {Stubbs}, {Szapudi},
  {Szkody}, {Szokoly}, {Tabachnik}, {Tsvetanov}, {Uomoto}, {Vogeley}, {Voges},
  {Waddell}, {Walterbos}, {Wang}, {Watanabe}, {Weinberg}, {White}, {White},
  {Wilhite}, {Wolfe}, {Yasuda}, {York}, {Zehavi}, \& {Zheng}}]{SDSS_EDR}
{Stoughton}, C., {Lupton}, R.~H., {Bernardi}, M., {et~al.} 2002, \aj, 123, 485

\bibitem[{{Strateva} {et~al.}(2001){Strateva}, {Ivezi{\'c}}, {Knapp},
  {Narayanan}, {Strauss}, {Gunn}, {Lupton}, {Schlegel}, {Bahcall}, {Brinkmann},
  {Brunner}, {Budav{\'a}ri}, {Csabai}, {Castander}, {Doi}, {Fukugita}, {Gy{\H
  o}ry}, {Hamabe}, {Hennessy}, {Ichikawa}, {Kunszt}, {Lamb}, {McKay},
  {Okamura}, {Racusin}, {Sekiguchi}, {Schneider}, {Shimasaku}, \&
  {York}}]{Strateva_et_al_2001}
{Strateva}, I., {Ivezi{\'c}}, {\v Z}., {Knapp}, G.~R., {et~al.} 2001, \aj, 122,
  1861

\bibitem[{{Strauss} {et~al.}(2002){Strauss}, {Weinberg}, {Lupton}, {Narayanan},
  {Annis}, {Bernardi}, {Blanton}, {Burles}, {Connolly}, {Dalcanton}, {Doi},
  {Eisenstein}, {Frieman}, {Fukugita}, {Gunn}, {Ivezi{\'c}}, {Kent}, {Kim},
  {Knapp}, {Kron}, {Munn}, {Newberg}, {Nichol}, {Okamura}, {Quinn}, {Richmond},
  {Schlegel}, {Shimasaku}, {SubbaRao}, {Szalay}, {Vanden Berk}, {Vogeley},
  {Yanny}, {Yasuda}, {York}, \& {Zehavi}}]{Strauss_et_al_2002}
{Strauss}, M.~A., {Weinberg}, D.~H., {Lupton}, R.~H., {et~al.} 2002, \aj, 124,
  1810

\bibitem[{{Treu} {et~al.}(2003){Treu}, {Ellis}, {Kneib}, {Dressler}, {Smail},
  {Czoske}, {Oemler}, \& {Natarajan}}]{Treu_et_al_2003}
{Treu}, T., {Ellis}, R.~S., {Kneib}, J., {et~al.} 2003, \apj, 591, 53

\bibitem[{Turlach(1993)}]{Turlach_Rev}
Turlach, B.~A. 1993, in CORE and Institut de Statistique

\bibitem[{{Wang} \& {Heckman}(1996)}]{Wang&Heckman_1996}
{Wang}, B. \& {Heckman}, T.~M. 1996, \apj, 457, 645

\end{thebibliography}

\end{document}